\documentclass[%
 amsmath,amssymb,
 aps, physrev,
pra,
twocolumn
]{revtex4-2}

\usepackage{graphicx}
\usepackage{dcolumn}
\usepackage{bm}

\begin{document}


\title{\textbf{
Free-Particle Green’s Function Matrix Elements over Spherical Gaussian and Plane-Wave–Modulated Gaussian Basis Functions} 
}%

\author{Dibyendu Mahato}
 \author{Wojciech Skomorowski}%
 \email{Contact author: w.skomorowski@cent.uw.edu.pl}
\affiliation{%
 Centre of New Technologies, University of Warsaw, Banacha 2c, 02-097 Warsaw, Poland.
}%

\date{\today}

\begin{abstract}
Free-particle Green’s function plays a central role in the theoretical description of electron scattering and autoionization processes in quantum physics and chemistry. Recently, Gaussian basis set approaches have become increasingly important in applications to unbound and metastable electronic states. However, the practical use of such methods has been limited by the lack of efficient and compact analytical expressions for matrix elements of the free-particle Green’s function in Gaussian-based representations. Here we present a novel, general analytical framework for the evaluation of one- and two-center matrix elements of the free-particle Green’s function over spherical Gaussian basis functions and plane-wave–modulated spherical Gaussians. The derivation is based on the Fourier transform of Gaussian functions together with the addition theorem of harmonic polynomials, leading to compact closed-form expressions and efficient recurrence relations. We also analyze the asymptotic behavior of the free-particle Green’s function matrix elements, which is essential in the description of low-energy continuum electrons using finite Gaussian basis sets.

\end{abstract}

\maketitle


\section{\label{sec:level1}Introduction}
Many important physical and chemical phenomena involve the interaction of matter with free electrons. These include electron--molecule scattering processes, photoionization, and autoionization processes such as Auger decay or Penning ionization. In all such cases, one or more electrons occupy continuum states, making the total electronic state non-square-integrable. Because of their continuum character, these states cannot be rigorously represented using finite \(L^2\)-integrable basis functions, as continuum wave functions do not satisfy bound-state boundary conditions and possess non-decaying asymptotic behavior. Nevertheless, quantum chemistry methods rely almost exclusively on finite \(L^2\) basis representations, and much of the success of modern electronic structure theory stems from the highly efficient evaluation of most common types of molecular integrals over Gaussian-type orbitals (GTOs), which traces back to the seminal work of Boys\cite{Boys1950}.

Applications of GTO-based methods to continuum-electron problems can be broadly divided into two categories. The first category includes non-Hermitian quantum chemistry approaches in which continuum states are approximated within an \(L^2\) framework using modified boundary conditions or complex-valued basis representation\cite{moiseyev2011non}. Examples include complex scaling, complex absorbing potentials, and complex basis function techniques, which are widely used for the description of metastable electronic states and the computation of resonance energies and lifetimes \cite{riss1993calculation,moiseyev1979autoionizing, mccurdy1978extension, moiseyev1998quantum, jagau2017extending}. The second category concerns scattering approaches in which continuum wave functions are expanded explicitly in a basis of Gaussian-type orbitals\cite{rescigno1974discrete, fliflet1978discrete, watson1979discrete, kaufmann1989universal}. This is often accompanied by nonlinear optimization of real or complex Gaussian functions in order to efficiently represent continuum states\cite{nestmann1990optimized, FAURE2002, Ammar2021b}. Such approaches are particularly attractive because they can be incorporated naturally into existing quantum chemistry software infrastructures that are heavily optimized for Gaussian integral evaluation. However, despite these advantages, applications of Gaussian basis functions to scattering and continuum problems remain considerably less developed than their bound-state counterparts.

One of the major challenges in Gaussian-based treatments of continuum processes is the efficient representation of the free-particle Green's operator
\begin{equation}
\hat{G}_0^{\pm}(E) = \lim_{\epsilon \to 0^{+}} \frac{1}{E - \hat{T} \pm i\epsilon },
\end{equation}
where \(\hat{T} \) is the particle's kinetic-energy operator. The free-particle Green's operator constitutes the central quantity in the Lippmann--Schwinger equation and determines the asymptotic structure of continuum scattering states\cite{Moyer1973, Greenman2017, Nestmann1990}. Consequently, an accurate and computationally efficient evaluation of matrix elements of \(\hat{G}_0^{\pm}(E)\) within a finite basis representation is essential for practical applications of GTOs to scattering and decay problems.

Several attempts have previously been made to derive analytical formulas for matrix elements of the free-particle Green's function over Cartesian Gaussian-type orbitals (CGTOs). Early work by Östlund \cite{Ostlund1975} introduced prototype integral formulas for \(s\)-type Cartesian Gaussians. These developments were later extended by Levin and co-workers \cite{Levin1978}, who derived analytical expressions up to \(f\)-type Cartesian Gaussians using partial-wave expansions of plane waves, although their formalism was restricted to axially symmetric systems. Subsequently, Čársky \textit{et al.} \cite{Carsky1996} developed general analytical expressions for arbitrary CGTO angular momenta. In their formulation, one-center integrals were evaluated through explicit radial integration combined with separate angular contributions, while two-center integrals were generated by differentiation of \(s\)-type Gaussian expressions with respect to Gaussian centers.

Although formally exact, these Cartesian Gaussian formulations become increasingly complicated for higher angular momenta and two-center integrals. In practice, the resulting expressions contain a large number of intermediate terms and do not efficiently reuse common quantities across symmetry-related integrals, which limits their computational efficiency and practical applicability. Perhaps for this reason, such approaches have not found widespread use in modern quantum chemistry calculations involving continuum states.

In the present work, we introduce an alternative and generalized analytical framework for evaluating matrix elements of the free-particle Green's function using spherical Gaussian-type orbitals (SGTOs). In SGTOs, spherical symmetry properties and geometric dependencies are incorporated naturally through angular momentum coupling. In addition, spherical Gaussian representations reduce the number of basis functions required for a given angular momentum. For angular momentum \(l\), the number of pure spherical harmonic functions \((2l+1)\) is smaller than or equal to the number of Cartesian Gaussian functions \((l+1)(l+2)/2\), resulting in improved computational efficiency.

Our derivation is based on the momentum-space representation of the Green's function operator together with the Fourier transform of spherical Gaussians and the addition theorem of harmonic polynomials. In momentum space, the angular and radial contributions separate naturally, similarly to the case of standard one-electron integrals such as kinetic-energy matrix elements. This allows many intermediate quantities to be reused efficiently and leads to compact analytical expressions together with simple recurrence relations.

Besides conventional SGTOs, we also consider Gaussian basis functions modulated by plane-wave factors (PW-SGTOs). Such basis functions have previously been proposed for the description of continuum and highly excited states because they incorporate oscillatory behavior directly into the basis representation \cite{FIORI2012,mccurdy1978extension,FAURE2002,bubin2020,AMMAR2021,Ammar2021b}. Standard Gaussian functions decay too rapidly in the asymptotic region and therefore require very large basis sets to reproduce continuum oscillations accurately. In contrast, plane-wave-modulated Gaussians introduce complex exponents that naturally account for oscillatory structure while preserving the advantageous analytical properties of Gaussian integrals.
In our derivations we exploit the fact that PW-SGTOs can be expressed as linear combinations of SGTOs with complex-valued Gaussian centers using the addition theorem of harmonic polynomials. As a consequence, the resulting Green's function matrix elements retain essentially the same analytical structure as ordinary SGTO integrals, differing only through the appearance of complex-valued Gaussian positions.

In this work, we present a unified and computationally efficient framework for evaluating matrix elements of the free-particle Green's function involving both SGTOs and PW-SGTOs. The derived expressions are significantly more compact and computationally efficient than previously reported CGTO-based formulations \cite{Carsky1996,Levin1978}, while retaining full generality with respect to angular momentum. 
The angular part of the derived integrals has been independently verified against standard kinetic-energy matrix elements implemented in the \textsc{Libqints} integral library, which is part of the \textsc{Q-Chem} quantum chemistry package \cite{qchemshort}. Additionally, the complete free-particle Green's function matrix elements have been benchmarked against the previously reported results of Čársky \textit{et al.} \cite{Carsky1996}.
We also investigate the asymptotic behavior of the free-particle Green's function matrix elements, which is essential for accurately describing low-energy continuum electrons within finite Gaussian basis representations. In particular, we identify specific asymptotic regimes in which simplified analytical forms can be applied. The resulting formalism provides an efficient route for the implementation of free-particle Green's function matrix elements in modern Gaussian-based electronic structure packages and is expected to facilitate future applications to scattering and decay processes involving continuum electronic states.

The remainder of the paper is organized as follows. In Section~\ref{sec:level-II}, we present a detailed derivation of the free-particle Green's function matrix elements in the basis of complex and real SGTOs. The derivation is formulated starting from a general one-electron operator in momentum representation, followed by the explicit construction of the radial recurrence relations required for the free-particle Green's function operator. In Section~\ref{sec:level-V}, the formalism is extended to plane-wave-modified SGTO basis functions. Section~\ref{sec:level-VIII} presents numerical illustrations of the matrix elements together with their asymptotic behavior. Finally, conclusions and remarks about specific applications of the present developments are given in Section~\ref{sec:level-IX}.

\section{\label{sec:level-II}Matrix Elements in Spherical Gaussian Basis Sets}

\subsection{Momentum-space representation of one-electron matrix elements}
We begin with the definition of a standard spherical Gaussian-type orbital (SGTO) in coordinate space,
\begin{equation}\label{eq3}
\phi_{lm}(\alpha,\mathbf{r})
=
N_l(\alpha)e^{-\alpha r^2}\mathcal{Y}_{lm}(\mathbf{r}),
\end{equation}
where
$\mathcal{Y}_{lm}(\mathbf{r})=r^lY_{lm}(\widehat{\mathbf{r}})$
is a solid harmonic \cite{louck1981angular}, and \(Y_{lm}(\widehat{\mathbf{r}})\) is the usual spherical harmonic, taken here with the Condon--Shortley phase convention \cite{Condon1935}. The normalization constant is \begin{equation} \label{eq5}
N_l(\alpha) =
\left[ \frac{2(2\alpha)^{l+3/2}}{\Gamma(l+3/2)} \right]^{1/2}.
\end{equation}
With the Fourier-transform convention used in this work, the corresponding momentum-space SGTO has the same angular dependence\cite{Kuang1997}, with the normalization factor given by
\begin{equation} \label{eq6}
\widetilde{N}_l(\alpha) = \frac{N_l(\alpha)}{(2\alpha)^{l+3/2}}.
\end{equation}

We consider matrix elements of a one-electron operator \(\widehat{\mathrm{O}}\) between two SGTOs centered at \(\mathbf{A}\) and \(\mathbf{B}\),
\begin{equation} \label{eq7}
O_{l_am_a}^{l_bm_b}(\alpha,\beta,\mathbf{A},\mathbf{B})
=\langle \phi_{l_am_a}(\alpha,\mathbf{r}-\mathbf{A})| \widehat{\mathrm{O}} | \phi_{l_bm_b}(\beta,\mathbf{r}-\mathbf{B}) \rangle .
\end{equation}
To transform this expression to momentum space, we insert the completeness relation
\begin{equation}
\hat{1}
= \int d\mathbf{p}\, |\mathbf{p}\rangle\langle\mathbf{p}|
\end{equation}
on both sides of the operator \(\widehat{\mathrm{O}}\). This gives
\begin{align}
&O_{l_am_a}^{l_bm_b}(\alpha,\beta,\mathbf{A},\mathbf{B})\nonumber
=&\\ \label{eq110}
&\int d\mathbf{p}\int d\mathbf{q}\,
\langle \phi_{l_am_a}(\alpha,\mathbf{r}-\mathbf{A}) |\mathbf{p}\rangle \langle\mathbf{p}|\widehat{\mathrm{O}}|\mathbf{q}\rangle
\langle\mathbf{q}|\phi_{l_bm_b}(\beta,\mathbf{r}-\mathbf{B})\rangle .
\end{align}
Here we focus on one-electron operators that are diagonal in momentum space, such as the outgoing free-particle Green's function in the \(k\)-representation (in atomic units):
\begin{equation}\label{eq11}
\langle \mathbf{p}|\widehat{\mathrm{G}}^{(+)}(k)|\mathbf{q} \rangle
= \lim_{\epsilon \to 0^{+}} \frac{1}{k^2-q^2+i\epsilon} \delta(\mathbf{p}-\mathbf{q}),
\end{equation}
or the kinetic-energy:
\begin{equation}\label{eq10}
\langle \mathbf{p}|\widehat{\mathrm{T}}|\mathbf{q} \rangle = \frac{q^2}{2}\delta(\mathbf{p}-\mathbf{q}).
\end{equation}
More generally, for a momentum-diagonal operator we write
\begin{equation}
\langle \mathbf{p}|\widehat{\mathrm{O}}|\mathbf{q} \rangle = \widehat{\mathrm{O}}(q)\delta(\mathbf{p}-\mathbf{q}).
\end{equation}

The momentum-space representations of SGTOs centered at \(\mathbf{A}\) and \(\mathbf{B}\) are
\begin{equation} \label{eq8}
\langle
\phi_{l_am_a}(\alpha,\mathbf{r}-\mathbf{A})|\mathbf{p}\rangle=i^{l_a}\widetilde{N}_{l_a}(\alpha)e^{-p^2/4\alpha} e^{i\mathbf{p}\cdot\mathbf{A}} \mathcal{Y}^{*}_{l_am_a}(\mathbf{p}),
\end{equation}
and
\begin{equation}\label{eq9}
\langle\mathbf{q}|\phi_{l_bm_b}(\beta,\mathbf{r}-\mathbf{B})\rangle
=(-i)^{l_b}\widetilde{N}_{l_b}(\beta) e^{-q^2/4\beta} e^{-i\mathbf{q}\cdot\mathbf{B}} \mathcal{Y}_{l_bm_b}(\mathbf{q}).
\end{equation}
Using Eqs.~\eqref{eq8} and \eqref{eq9}, together with the diagonal momentum-space representation of \(\widehat{\mathrm{O}}\), the matrix element in Eq.~\eqref{eq110} becomes
\begin{align}    
&O_{l_am_a}^{l_bm_b}(\alpha,\beta,\mathbf{R})
=\nonumber \label{eq12}
\\&C_{l_al_b}(\alpha,\beta)\int d\mathbf{q}\,\widehat{\mathrm{O}}(q)e^{i\mathbf{q}\cdot\mathbf{R}}e^{-\eta q^2}\mathcal{Y}^{*}_{l_am_a}(\mathbf{q}) \mathcal{Y}_{l_bm_b}(\mathbf{q}),
\end{align}
where
\begin{equation}\label{eq13}
C_{l_al_b}(\alpha,\beta)=i^{l_a}(-i)^{l_b}\widetilde{N}_{l_a}(\alpha)\widetilde{N}_{l_b}(\beta),
\end{equation}
\begin{equation}\label{eq14}
\eta = \frac{\alpha+\beta}{4\alpha\beta},
\end{equation}
and
\begin{equation}\label{eq15}
\mathbf{R} = \mathbf{A}-\mathbf{B}.
\end{equation}

The next step is to express the product of two solid harmonics in terms of coupled spherical harmonics and Gaunt coefficients $\langle lm | l_am_a|l_bm_b \rangle$. We also use the partial-wave expansion of the plane-wave factor,
\begin{equation}\label{eq23}
e^{i\mathbf{q}\cdot\mathbf{R}} =
4\pi \sum_{lm} i^{l} j_l(qR) Y^{*}_{lm}(\widehat{\mathbf{q}}) Y_{lm}(\widehat{\mathbf{R}}),
\end{equation}
where \(j_L(qR)\) is the spherical Bessel function and \(R=|\mathbf{R}|\).

After carrying out the angular integration using the orthogonality of spherical harmonics, the general expression for the matrix element can be written as
\begin{align}
O_{l_am_a}^{l_bm_b}(\alpha,\beta,\mathbf{R})
&=
4\pi C_{l_al_b}(\alpha,\beta) \nonumber\\ \times \label{eq25}
&\sum_{lm} i^l \langle lm | l_am_a|l_bm_b \rangle Y_{lm}(\widehat{\mathbf{R}}) I_l(\eta,R),
\end{align}
where the radial momentum integral is defined as
\begin{equation}\label{eq26}
I_l(\eta,R) = \int_{0}^{\infty} \widehat{\mathrm{O}}(q) q^{2+l_a+l_b} e^{-\eta q^2} j_l(qR) \,dq .
\end{equation}

In standard applications of quantum chemistry, it is customary to employ real spherical harmonics \(X_{l}^{\mu}(\widehat{\mathbf{R}})\) rather than complex-valued spherical harmonics. The real spherical harmonics \(X_{l}^{\mu}(\widehat{\mathbf{R}})\) are defined as
\begin{equation}\label{eq28}
X_l^\mu(\widehat{\mathbf{R}}) = 
\begin{cases} 
\frac{1}{\sqrt{2}}\Big(Y_{l,-|\mu|}(\widehat{\mathbf{R}})+(-1)^\mu Y_{l,|\mu|}(\widehat{\mathbf{R}})\Big) & \text{for } \mu > 0 \\
Y_{l,0}(\widehat{\mathbf{R}}) & \text{for } \mu=0 \\
\frac{i}{\sqrt{2}}\Big(Y_{l,-|\mu|}(\widehat{\mathbf{R}})-(-1)^\mu Y_{l,|\mu|}(\widehat{\mathbf{R}})\Big) & \text{for } \mu < 0. 
\end{cases}
\end{equation}
Here the indices \(l\) and \(m\) refer to the usual complex spherical harmonics, while \(l\) and \(\mu\) denote the corresponding real spherical harmonics (in the present work the index \(\mu\) is used whenever real spherical harmonics are employed).  
The use of real spherical harmonics leads to the same general expression for the matrix elements \(O_{l_a\mu_a}^{l_b\mu_b}(\alpha,\beta,\mathbf{R})\):
\begin{align}
O_{l_a\mu_a}^{l_b\mu_b}(\alpha,\beta,\mathbf{R})
&= 4\pi C_{l_al_b}(\alpha,\beta) \nonumber
\\ \label{eq27} &\times \sum_{l\mu} i^l \langle l\mu | l_a\mu_a | l_b\mu_b \rangle X_l^{\mu}(\widehat{\mathbf{R}}) I_l(\eta,R),
\end{align}
except that the standard Gaunt coefficients are replaced by the corresponding real spherical harmonic Gaunt coefficients
$\langle l\mu | l_a\mu_a | l_b\mu_b \rangle$. Specific formulas and properties for both complex and real Gaunt coefficients are provided in Appendix~\ref{appG}. The expression in Eq.~\eqref{eq27} is fully general and can be applied to any one-electron operator that is diagonal in momentum space\cite{gao2011gen1int}.

It is also worth noting that the one-center integrals, corresponding to the case \(R=0\), simplify considerably and become diagonal with respect to the angular momentum quantum numbers. Using the orthogonality properties of spherical harmonics together with the relation \(j_l(0)=\delta_{l0}\), the general expression of Eq.~\eqref{eq25} reduces to
\begin{equation}
\label{eq35}
O_{l_am_a}^{l_bm_b}(\alpha,\beta) =C_{l_al_b}(\alpha,\beta) I_0(\eta,0)\delta_{l_al_b}\,\delta_{m_am_b}.
\end{equation}
 
\subsection{Recursive evaluation of two-center free-particle Green’s function integrals}
We now consider the specific case of the free-particle Green's function. In the present work, we focus only on the outgoing Green's operator \(\widehat{\mathrm{G}}^{(+)}(k_0)\) defined in Eq.~\eqref{eq11}, which is the form typically employed in quantum chemistry applications (see e.g. Ref.~\cite{Carsky1996}). The matrix elements of the incoming Green's operator \(\widehat{\mathrm{G}}^{(-)}(k_0)\) can be obtained directly by Hermitian conjugation of \(\widehat{\mathrm{G}}^{(+)}(k_0)\).

By virtue of the general expression in Eq.~\eqref{eq27}, the matrix elements of the Green's function in the basis of complex spherical harmonics for an electron with energy \(E_0=k_0^2/2\) can be written as
\begin{align}
&G_{l_am_a}^{l_bm_b}(\alpha,\beta,\mathbf{R},k_0)=4\pi C_{l_al_b}(\alpha,\beta)
\nonumber\\ \label{eq50}
&\times\sum_{l} i^l\langle{lm}|l_am_a|{l_bm_b}\rangle Y_{lm}(\widehat{\mathbf{R}})I_l^{P^{\prime}}(\eta,R,k_0),
\end{align}
or in the basis of real spherical harmonics
\begin{align}
&G_{l_a\mu_a}^{l_b\mu_b}(\alpha,\beta,\mathbf{R},k_0)
= 4\pi C_{l_al_b}(\alpha,\beta)
\nonumber
\\ \label{eq51} &\times \sum_{l\mu} i^l \langle l\mu|l_a\mu_a|l_b\mu_b\rangle
X_{l}^{\mu}(\widehat{\mathbf{R}}) I_l^{P^{\prime}}(\eta,R,k_0),
\end{align}
where the radial momentum integral is defined as
\begin{equation}\label{eq52}
I_l^{P^{\prime}}(\eta,R,k_0)=\int_{0}^{\infty}\frac{q^{P^{\prime}}e^{-\eta q^2}j_l(qR)}{k_0^2-q^2+i\epsilon}\,dq,
\end{equation}
with $P^{\prime}=2+l_a+l_b$. Integrals of this type can be evaluated efficiently using recurrence relations of the spherical Bessel functions. In particular, the following relation holds:
\begin{align}
I_{l+1}^{P^{\prime}+1}&(\eta,R, k_0) =
\nonumber
\\ \label{eq40} &\frac{2l+1}{R}I_l^{P^{\prime}}(\eta,R, k_0) - I_{l-1}^{P^{\prime}+1}(\eta,R, k_0),
\quad l\ge 1.
\end{align}
Therefore, it is sufficient to evaluate only the integrals corresponding to \(l=0\) and \(l=1\), while all higher-order terms can be obtained recursively from Eq.~\eqref{eq40}.

The required radial integrals may further be generated by differentiation with respect to \(\eta\) and \(R\). In practice, it is sufficient to evaluate the fundamental integrals \(I_0^2\) and its first derivative with respect to \(R\), namely \(I_1^3\). All higher-order terms can then be obtained through differentiation with respect to \(\eta\). The general recurrence formula for the Green's function integrals can thus be written as


\begin{equation}\label{eq53}
I_l^{2+2n+l}(\eta,R,k_0) = \frac{\pi}{4 R}\sum_{q=0}^n \binom{n}{q} (-1)^{q} k_0^{2(n-q)} g_l^{(q)}(\eta,R,k_0).
\end{equation}

The Eq.~\eqref{eq53} will generate all the required integrals for \(I_0^2\), \(I_0^4\), \(I_0^6\), ... for \(l=0\) and \(I_1^3\), \(I_1^5\), \(I_1^7\), ... for \(l=1\) respectively. Whereas, the remaining higher-order terms \(I_2^4\), \(I_2^6\), \(I_3^5\) and others can be obtained by using the Eq.~\eqref{eq40}. The initial terms entering Eq.~\eqref{eq53} are
\begin{equation}\label{eq55}
g^{(0)}_0(\eta,R,k_0) = g^{(0)}_{0-}(\eta,R,k_0),
\end{equation}
and
\begin{equation}\label{eq56}
g^{(0)}_1(\eta,R,k_0) = \frac{1}{R}g^{(0)}_{0-}(\eta,R,k_0) + ik_0g^{(0)}_{0+}(\eta,R,k_0) + \frac{2e^{-\frac{R^2}{4\eta}}}{\sqrt{\eta\pi}}.
\end{equation}
Here \(g^{(0)}_{0-}(\eta,R,k_0)\) and \(g^{(0)}_{0+}(\eta,R,k_0)\)
are defined in terms of the complementary error function \(\mathrm{Erfc}\):
\begin{align}
g^{(0)}_{0\pm}(\eta,R,k_0) =& e^{-\eta k_0^2-ik_0 R}
\mathrm{Erfc} \left[z_{+}(\eta,R,k_0)\right]
\nonumber \\ \label{eq57} &\pm 
e^{-\eta k_0^2+ik_0 R}\mathrm{Erfc}\left[z_{-}(\eta,R,k_0)\right],
\end{align}
where
\begin{equation}\label{eqd5}
    z_{\pm}(\eta,R,k_0)=-i\Big(\eta^{1/2}k_0\pm\frac{iR}{2\eta^{1/2}}\Big).
\end{equation}
The higher-order terms of \(g^{(n)}_{0\pm}(\eta,R,k_0)\) for \(n\ge1\) are given by
\begin{equation}\label{eq59}
g^{(n)}_{0+}(\eta,R,k_0) = \frac{2ik_0}{\sqrt{\pi}}
\sum_{k=0}^{n-1} \binom{n-1}{k} h^{(n-k-1)}(\eta,R,k_0) q^{(k)}_{1/2}(\eta),
\end{equation}
and
\begin{equation}\label{eq60}
g^{(n)}_{0-}(\eta,R,k_0)
= \frac{R}{\sqrt{\pi}} \sum_{k=0}^{n-1} \binom{n-1}{k} h^{(n-k-1)}(\eta,R,k_0) q^{(k)}_{3/2}(\eta),
\end{equation}
where
\begin{equation}\label{eq61}
q^{(n)}_{m/2}(\eta) = (-1)^n
\frac{ \Gamma\left(\frac{m}{2}+n\right)}{\Gamma\left(\frac{m}{2}\right)}
\frac{1}{\eta^{\frac{m}{2}+n}},
\end{equation}
and
\begin{align}
h^{(n)}&(\eta,R,k_0) = \nonumber
\\ \label{eq62} & \sum_{k=0}^{n-1} \binom{n-1}{k}h^{(n-k-1)}(\eta,R,k_0)s^{(k)}(\eta,R,k_0).
\end{align}
The initial value of \(h^{(n)}(\eta,R,k_0)\) is
\begin{equation}\label{eq63}
h^{(0)}(\eta,R,k_0) = e^{-\frac{R^2}{4\eta}},
\end{equation}
while the \(n^{\mathrm{th}}\)-order terms of \(s(\eta,R,k_0)\) are
\begin{equation}\label{eq64}
s^{(n)}(\eta,R,k_0)=
\begin{cases}
k_0^2+\dfrac{R^2}{4\eta^2},
& n=0,
\\[0.4cm]
(-1)^n
\left(\dfrac{R^2}{4}\right)
\dfrac{(n+1)!}{\eta^{n+2}},
& n\ge1.
\end{cases}
\end{equation}
After obtaining \(g^{(n)}_{0-}(\eta,R,k_0)\) and \(g^{(n)}_{0+}(\eta,R,k_0)\), the higher-order terms of \(g_1^{(n)}(\eta,R,k_0)\) can be generated straightforwardly from Eq.~\eqref{eq56},
\begin{align}
    \label{eq65}
g^{(n)}_1(\eta,R,k_0) =
\frac{1}{R}g^{(n)}_{0-}(\eta,R,k_0)
+ ik_0g^{(n)}_{0+}(\eta,R,k_0)
\nonumber\\
+ \frac{2}{\sqrt{\pi}} \sum_{k=0}^{n}\binom{n}{k} h^{(n-k)}(\eta,R,k_0) q^{(k)}_{1/2}(\eta).
\end{align}

\subsection{One-center matrix elements}
In the special case of one-center matrix elements, corresponding to \(R=0\), the general expression in Eq.~\eqref{eq35} leads to a particularly simple form for the free-particle Green's function matrix elements:
\begin{equation}\label{eq67}
G_{l_am_a}^{l_bm_b}(\alpha,\beta,k_0) = C_{l_al_b}(\alpha,\beta) J^{2Q}(\eta,k_0)\delta_{l_al_b}\,\delta_{m_am_b},
\end{equation}
where the radial integral entering Eq.~\eqref{eq67} is defined as
\begin{equation}\label{eq68}
J^{2Q}(\eta,k_0) = \int_{0}^{\infty} \frac{ q^{2Q} e^{-\eta q^2}}{k_0^2-q^2+i\epsilon}\,dq,
\end{equation}
with  $2Q=2+l_a+l_b$. Because of the condition \(l_a=l_b\), the quantity \(Q\) is always a positive integer.

The general recurrence formula for evaluating \(J^{2Q}(\eta,k_0)\) is
\begin{equation}\label{eq69}
J^{2Q}(\eta,k_0) = k_0^{2Q}J_0(\eta,k_0) - \sum_{m=0}^{Q-1} k_0^{2(Q-m-1)} I_{2m}(\eta), \; Q\ge1,
\end{equation}
where \(J_0(\eta,k_0)\) can be expressed in terms of the Faddeeva function $W(z)=e^{-z^2}\mathrm{Erfc}(-iz)$
as
\begin{equation}\label{eq70}
J_0(\eta,k_0) = -\frac{\pi i}{2k_0} W(\sqrt{\eta}k_0),
\end{equation}
and
\begin{equation}\label{eq71}
I_{2m}(\eta) = \sqrt{\pi} \frac{(2m-1)!!}{2^{m+1}\eta^{(2m+1)/2}}, \qquad m\ge0.
\end{equation}

Overall, the recurrence relations given above for both one-center and two-center matrix elements of the free-particle Green's function, together with their asymptotic forms discussed in the next subsection, provide a compact and efficient framework for evaluating matrix elements up to arbitrary angular momentum. The resulting expressions are significantly simpler and computationally more efficient than previously reported formulations based on Cartesian Gaussian basis sets.

In addition to the free-particle Green's function operator, Appendix~\ref{appK} also presents analogous recurrence formulas for kinetic-energy matrix elements. Although the kinetic-energy operator itself is considerably simpler, it provides a useful reference system for validating the angular part of the present formalism, since both operators possess the same angular and geometric structure.

\subsection{Asymptotic limits of the radial integrals}

The radial integrals reported above involve the complementary error function \(\mathrm{Erfc}\), and therefore their numerical stability must be analyzed carefully over different parameter regimes.
For most Gaussian exponents, the direct evaluation of the SGTO Green's function matrix elements using Eq.~\eqref{eq57} is numerically stable. However, two limiting regimes require special consideration due to the asymptotic behavior of the complementary error function. These limits can be identified from the definition of the functions \(g^{(0)}_{0\pm}(\eta,R,k_0)\) in Eq.~\eqref{eq57} and the corresponding arguments \(z_{\pm}\) defined in Eq.~\eqref{eqd5}.

The first limiting regime corresponds to a large real part of \(z_{\pm}\), i.e. $R / \sqrt{\eta} \gg 1$,
where the interatomic separation becomes large compared to the effective Gaussian width. Using the asymptotic expansion of the complementary error function
it can be shown that in this case the radial integrals in Eq.~\eqref{eq53} asymptotically reduce to
\begin{equation}\label{eq66}
I_l^{2+2n+l}(\eta,R,k_0)
=
-\frac{\pi}{2R}
k_0^{2n}
e^{-\eta k_0^2}
e^{ik_0R}
\left(
\frac{1}{R}-ik_0
\right)^l .
\end{equation}
Here \(l=0\) corresponds to the integrals \(I_0^{2+2n}(\eta,R,k_0)\), while \(l=1\) corresponds to \(I_1^{2+2n+1}(\eta,R,k_0)\). This behavior follows from the fact that for large \(R/\sqrt{\eta}\) one of the complementary error functions approaches the constant value \(2\), while the remaining contributions in \(g^{(0)}_{0\pm}\) become exponentially suppressed. The remaining higher-order integrals can subsequently be generated using the general recurrence relation given in Eq.~\eqref{eq40}. Therefore, this asymptotic regime does not lead to any numerical instabilities.

A more complicated situation arises when the imaginary part of \(z_{\pm}\) becomes large, i.e., $\sqrt{\eta}k_0 \gg 1$. This regime occurs for very diffuse Gaussian functions, corresponding to small Gaussian exponents \(\alpha\) and \(\beta\). In this case, direct evaluation of Eq.~\eqref{eq57} may lead to numerical instabilities due to cancellation between exponentially large and exponentially small terms. To avoid this problem, the asymptotic expansion of the complementary error function for large complex arguments must be applied directly. The corresponding expressions for \(g^{(0)}_{0\pm}(\eta,R,k_0)\) then become
\begin{align}
    g_{0\pm}^{(0)}(\eta,R,k_0)=&\frac{e^{-\frac{R^2}{4\eta}}}{\sqrt{\pi}z_{+}}\sum_{n=0}^{q} \frac{(-1)^n (2n - 1)!!}{(2z_{+}^2)^n} \nonumber
    \\ \label{eqd19}&\pm\frac{e^{-\frac{R^2}{4\eta}}}{\sqrt{\pi}z_{-}}\sum_{n=0}^{q} \frac{(-1)^n (2n - 1)!!}{(2z_{-}^2)^n}.
\end{align}
In this regime, Eq.~\eqref{eqd19} replaces the direct evaluation of Eq.~\eqref{eq57}, while all remaining recurrence relations follow identically from Eqs.~\eqref{eq40}--\eqref{eq65}.

A similar asymptotic treatment can also be applied to the one-center case. By expanding the complementary error function entering Eq.~\eqref{eq70}, one obtains the asymptotic expression for $\sqrt{\eta}k_0 \gg 1$:
\begin{equation}\label{eq101}
J_0(\eta,k_0)=
\frac{\sqrt{\pi}}{2\eta^{1/2}k_0^2}
\sum_{n=0}^{q}
\frac{(2n-1)!!}{\left(2\eta k_0^2\right)^n}.
\end{equation}

It is worth noting that, in this asymptotic regime, both the radial integrals and the corresponding Green's function matrix elements become purely real in the case of real SGTOs. For the one-center case, this follows directly from Eq.~\eqref{eq101}. For two-center matrix elements, this behavior can be understood from the asymptotic expressions for \(g_{0\pm}^{(0)}\) given in Eq.~\eqref{eqd19}, together with the prefactors appearing in Eq.~\eqref{eq56} for the function \(g^{(0)}_1\).

\section{\label{sec:level-V}Matrix Elements over Plane-Wave-Modulated Spherical Gaussian Functions}
We now consider matrix elements in the basis of spherical Gaussian functions modulated by a plane-wave factor (PW-SGTOs). Such functions have found several applications in quantum chemistry, including calculations employing periodic boundary conditions and gauge-origin-independent treatments of molecules in external magnetic fields. In the present context, these functions are particularly attractive for the description of continuum electrons, since the plane-wave factor naturally incorporates oscillatory behavior and may significantly reduce the number of Gaussian functions required to represent continuum-like states.

The evaluation of free-particle Green's function matrix elements over plane-wave-modulated Gaussian functions has received relatively little attention. Colle \textit{et al.}~\cite{colle1988} derived analytical prototype integrals involving Cartesian Gaussians and plane waves. However, their formulation contained inconsistencies in the overlap normalization factors, which complicates its practical application. Here we derive the corresponding matrix elements using the addition theorem applied to both complex and real spherical harmonics.

The key observation underlying the present formulation is that the plane-wave factor can be absorbed into a shift of the Gaussian center into the complex plane\cite{allison1973, dose1974}. This property is particularly simple in the case of \(s\)-type Gaussian functions:
\begin{equation}\label{eq72}
e^{i\mathbf{k}\cdot(\mathbf{r}-\mathbf{C})}
e^{-\alpha (\mathbf{r}-\mathbf{C})^2}
=
\mathcal{N}_{\mathbf{k}}(\alpha)
e^{-\alpha (\mathbf{r}-\mathbf{C}^{\dagger})^2},
\end{equation}
where
\begin{equation}
\mathbf{C}^{\dagger}
=
\mathbf{C}
+
\frac{i\mathbf{k}}{2\alpha},
\end{equation}
and
\begin{equation}
\mathcal{N}_{\mathbf{k}}(\alpha)
=
e^{-\frac{\mathbf{k}^2}{4\alpha}}.
\end{equation}

To proceed further, we expand the plane-wave-modulated spherical harmonics using the addition theorem for solid spherical harmonics,

\begin{widetext}
\begin{equation}\label{eq75}
e^{i\mathbf{k}\cdot(\mathbf{r}-\mathbf{C})}
\phi_{lm}(\alpha,\mathbf{r}-\mathbf{C})
=
4\pi
N_l(\alpha)
\mathcal{N}_{\mathbf{k}}(\alpha)
\sum_{l^{\prime}m^{\prime}}
G(lm|l^{\prime}m^{\prime})
\frac{
\phi_{l^{\prime}m^{\prime}}(\alpha,\mathbf{r}-\mathbf{C}^{\dagger})
}{
N_{l^{\prime}}(\alpha)
}
\mathcal{Y}_{l-l^{\prime},m-m^{\prime}}
\Big(
\frac{i\mathbf{k}}{2\alpha}
\Big),
\end{equation}
\end{widetext}
where the coefficients \(G(lm|l^{\prime}m^{\prime})\) are given explicitly in Appendix~C. An analogous expression can also be written in momentum space by Fourier transforming the spherical Gaussian function
\(\phi_{l^{\prime}m^{\prime}}(\alpha,\mathbf{r}-\mathbf{C}^{\dagger})\).
Based on this observation, and using Eq.~\eqref{eq75} together with the Fourier transforms of
\(\phi_{l_am_a}(\alpha,\mathbf{r}-\mathbf{A}^{\dagger})\)
and
\(\phi_{l_bm_b}(\beta,\mathbf{r}-\mathbf{B}^{\dagger})\),
the general expression for matrix elements in the PW-SGTO basis can be written as
\begin{widetext}
\begin{eqnarray}\label{eq77}
O_{l_am_a}^{l_bm_b}(\alpha,\beta,\mathbf{R}^{\dagger},\mathbf{k}_1,\mathbf{k}_2)
&=&
(4\pi)^2
N_{l_al_b}(\alpha,\beta)
[\mathcal{N}_{\mathbf{k}_1}(\alpha)]^{*}
\mathcal{N}_{\mathbf{k}_2}(\beta)
\sum_{l_a^{\prime}m_a^{\prime}}
\sum_{l_b^{\prime}m_b^{\prime}}
G(l_am_a|l_a^{\prime}m_a^{\prime})
G(l_bm_b|l_b^{\prime}m_b^{\prime})
\nonumber\\
&&\times
\frac{
O_{l_a^{\prime}m_a^{\prime}}^{l_b^{\prime}m_b^{\prime}}
(\alpha,\beta,\mathbf{R}^{\dagger})
}{
N_{l_a^{\prime}l_b^{\prime}}(\alpha,\beta)
}
\mathcal{Y}^{*}_{l_a-l_a^{\prime},m_a-m_a^{\prime}}
\Big(
\frac{i\mathbf{k}_1}{2\alpha}
\Big)
\mathcal{Y}_{l_b-l_b^{\prime},m_b-m_b^{\prime}}
\Big(
\frac{i\mathbf{k}_2}{2\beta}
\Big),
\end{eqnarray}
\end{widetext}
where, for an arbitrary complex vector \(\mathbf{C}\),
\begin{equation}
[\mathcal{Y}_{lm}(\mathbf{C})]^{*}
=
(-1)^m
\mathcal{Y}_{l,-m}(\mathbf{C}^{*}).
\end{equation}
In Eq.~\eqref{eq77}, the normalization constant \(N_{l_al_b}(\alpha,\beta)\) and the complex displacement vector \(\mathbf{R}^{\dagger}\) are defined as
\begin{equation}\label{eq78}
N_{l_al_b}(\alpha,\beta) = N_{l_a}(\alpha)N_{l_b}(\beta),
\end{equation}
and
\begin{equation}\label{eq79}
\mathbf{R}^{\dagger} = (\mathbf{A}-\mathbf{B}) - i\left(\frac{\mathbf{k}_1}{2\alpha} + \frac{\mathbf{k}_2}{2\beta} \right).
\end{equation}
The matrix elements
\( O_{l_a^{\prime}m_a^{\prime}}^{l_b^{\prime}m_b^{\prime}}(\alpha,\beta,\mathbf{R}^{\dagger})\)
appearing in Eq.~\eqref{eq77} have exactly the same form as in Eq.~\eqref{eq25} for standard SGTOs, except that they now depend on the complex vector \(\mathbf{R}^{\dagger}\) with $R^{\dagger} = |\mathbf{R}^{\dagger}|$. Recurrence relations for the evaluation of solid spherical harmonics with complex arguments $\mathcal{Y}_{lm}(\mathbf{C})$ are discussed in Appendix~\ref{appB}.

We now derive the analogous expressions in the basis of real spherical harmonics. To this end, the addition theorem for real solid spherical harmonics must be employed. In this case, however, the derivation becomes somewhat more involved than for complex spherical harmonics, since the decomposition of the phase factor into cosine- and sine-like components introduces additional terms in the summation. Using the addition theorem for real regular solid harmonics, the plane-wave-modulated real SGTO can be expanded as \cite{rico2013}:
\begin{widetext}
\begin{equation}\label{eq87}
e^{i\mathbf{k}\cdot(\mathbf{r}-\mathbf{C})}
\phi_{l\mu}(\alpha,\mathbf{r}-\mathbf{C})
=
N_{l}(\alpha)
N_{l\mu}^{\prime}
\mathcal{N}_{\mathbf{k}}(\alpha)
\sum_{l^{\prime}=0}^{l}
\sum_{\mu^{\prime}=-l^{\prime}}^{l^{\prime}}
\frac{
\phi_{l^{\prime}\mu^{\prime}}(\alpha,\mathbf{r}-\mathbf{C}^{\dagger})
}{
N_{l^{\prime}}(\alpha)
N_{l^{\prime}\mu^{\prime}}^{\prime}
}
\sum_{\mu^{\prime\prime}}
z_{l-l^{\prime}}^{\mu^{\prime\prime}}
\Big(
\frac{i\mathbf{k}}{2\alpha}
\Big)
c_{l^{\prime}\mu^{\prime}\mu^{\prime\prime}}^{l\mu}.
\end{equation}
\end{widetext}
Here
\(z_l^{\mu}(\mathbf{r})\)
denotes the unnormalized real spherical harmonic:
\begin{equation}\label{eq85}
z_l^{\mu}(\mathbf{r})
=
(N^\prime_{l\mu})^{-1}
X_l^{\mu}(\mathbf{r}),
\end{equation}
with auxiliary constant
\begin{equation}\label{eq86}
N^{\prime}_{l\mu}
=
\Bigg[
\frac{(2l+1)}{2\pi(1+\delta_{0,\mu})}
\frac{(l-|\mu|)!}{(l+|\mu|)!}
\Bigg]^{1/2}.
\end{equation}
The explicit evaluation of the PW-SGTO expansion in Eq.~\eqref{eq87} requires separate consideration of the cases \(\mu\ge0\) and \(\mu<0\). The corresponding general coefficients \(c_{l^{\prime}\mu^{\prime}\mu^{\prime\prime}}^{l\mu}\) for each case can be derived from Eqs.~\eqref{eqa6} and \eqref{eqa7} in Appendix~\ref{appA}.

Further, the general compact expression for matrix elements in the basis of plane-wave-modulated real SGTOs becomes
\begin{widetext}
\begin{eqnarray}\label{eq88}
O_{l_a\mu_a}^{l_b\mu_b}
(\alpha,\beta,\mathbf{R}^{\dagger},\mathbf{k}_1,\mathbf{k}_2)
&=&
D_{l_a\mu_a}^{l_b\mu_b}
(\alpha,\beta,\mathbf{k}_1,\mathbf{k}_2)
\sum_{l_a^{\prime}\mu_a^{\prime}}
\sum_{l_b^{\prime}\mu_b^{\prime}}
\frac{
O_{l_a^{\prime}\mu_a^{\prime}}^{l_b^{\prime}\mu_b^{\prime}}
(\alpha,\beta,\mathbf{R}^{\dagger})
}{
N_{l_a^{\prime}l_b^{\prime}}(\alpha,\beta)
N_{l_a^{\prime}\mu_a^{\prime}}^{\prime}
N_{l_b^{\prime}\mu_b^{\prime}}^{\prime}
}
\nonumber\\
&&\times
\sum_{\mu_a^{\prime\prime}\mu_b^{\prime\prime}}
c_{l_a^{\prime}\mu_a^{\prime}\mu_a^{\prime\prime}}^{l_a\mu_a}
c_{l_b^{\prime}\mu_b^{\prime}\mu_b^{\prime\prime}}^{l_b\mu_b}
z_{l_a-l_a^{\prime}}^{\mu_a^{\prime\prime}}
\Big(
-\frac{i\mathbf{k}_1}{2\alpha}
\Big)
z_{l_b-l_b^{\prime}}^{\mu_b^{\prime\prime}}
\Big(
\frac{i\mathbf{k}_2}{2\beta}
\Big),
\end{eqnarray}
\end{widetext}
where the combined normalization factor
\(D_{l_a\mu_a}^{l_b\mu_b}(\alpha,\beta,\mathbf{k}_1,\mathbf{k}_2)\)
is defined as
\begin{align}
   D_{l_a\mu_a}^{l_b\mu_b}
&(\alpha,\beta,\mathbf{k}_1,\mathbf{k}_2)
= \nonumber\\&  \label{eq90}
N_{l_al_b}(\alpha,\beta)
N_{l_a\mu_a}^{\prime}
N_{l_b\mu_b}^{\prime}
[\mathcal{N}_{\mathbf{k}_1}(\alpha)]^{*}
\mathcal{N}_{\mathbf{k}_2}(\beta).
\end{align}
The matrix elements 
\(
O_{l_a^{\prime}\mu_a^{\prime}}^{l_b^{\prime}\mu_b^{\prime}}
(\alpha,\beta,\mathbf{R}^{\dagger})
\)
appearing in Eq.~\eqref{eq88} have exactly the same form as the standard SGTO matrix elements in Eq.~\eqref{eq27}, except that they now depend on the complex displacement vector \(\mathbf{R}^{\dagger}\). 

Inspection of Eqs.~\eqref{eq77} and \eqref{eq88} shows that the introduction of the plane-wave modulation leads to additional angular-coupling coefficients and summations over auxiliary angular momentum quantum numbers arising from the addition theorem of solid harmonics. Although the resulting expressions may appear considerably more complicated than in the standard SGTO case, the additional summations contain only a relatively small number of terms in practical applications, as discussed in more detail in Appendix~\ref{appA}.
Specifically, depending on the signs of the two real spherical harmonic indices, i.e., whether \(\mu\ge0\) or \(\mu<0\), and similarly for \(\mu^{\prime}\), four different expressions arise for the evaluation of the corresponding matrix element in Eq.~\eqref{eq88}. These four possible combinations are considered explicitly in Eqs.~\eqref{eqa8}--\eqref{eqa11} of Appendix~\ref{appA}.

The generalized PW-SGTO formalism derived above is applicable to any one-electron operator diagonal in momentum space, including the free-particle Green's function and kinetic-energy operators. Here we present only the final expression for the free-particle Green's function matrix elements in the basis of PW-SGTOs. For complex spherical harmonics, Eq.~\eqref{eq77} leads to
\begin{widetext}
\begin{eqnarray}\label{eq100}
G_{l_am_a}^{l_bm_b}
(\alpha,\beta,\mathbf{R}^{\dagger},\mathbf{k}_1,\mathbf{k}_2,k_0)
&=&
(4\pi)^2
N_{l_al_b}(\alpha,\beta) [\mathcal{N}_{\mathbf{k}_1}(\alpha)]^{*}
\mathcal{N}_{\mathbf{k}_2}(\beta)
\sum_{l_a^{\prime}m_a^{\prime}}
\sum_{l_b^{\prime}m_b^{\prime}}
G(l_am_a|l_a^{\prime}m_a^{\prime})
G(l_bm_b|l_b^{\prime}m_b^{\prime})
\nonumber\\
&&\times
\frac{
G_{l_a^{\prime}m_a^{\prime}}^{l_b^{\prime}m_b^{\prime}}
(\alpha,\beta,\mathbf{R}^{\dagger},k_0)
}{
N_{l_a^{\prime}l_b^{\prime}}(\alpha,\beta)
}
\mathcal{Y}^{*}_{l_a-l_a^{\prime},m_a-m_a^{\prime}}
\Big(
\frac{i\mathbf{k}_1}{2\alpha}
\Big)
\mathcal{Y}_{l_b-l_b^{\prime},m_b-m_b^{\prime}}
\Big(
\frac{i\mathbf{k}_2}{2\beta}
\Big).
\end{eqnarray}
\end{widetext}
Here
\(
G_{l_a^{\prime}m_a^{\prime}}^{l_b^{\prime}m_b^{\prime}}
(\alpha,\beta,\mathbf{R}^{\dagger},k_0)
\)
are the standard SGTO Green's function matrix elements defined in Eq.~\eqref{eq51}, again evaluated for the complex vector \(\mathbf{R}^{\dagger}\). An analogous expression for real spherical harmonics follows directly by inserting the matrix elements
\(G_{l_a\mu_a}^{l_b\mu_b}(\alpha,\beta,\mathbf{R}^{\dagger},k_0)\)
from Eq.~\eqref{eq51} into the general formula of Eq.~\eqref{eq88}.

In a completely analogous manner, the corresponding expressions for kinetic-energy matrix elements can be obtained by substituting the matrix elements from Eqs.~\eqref{eq37} and \eqref{eq39}. These expressions are particularly useful for debugging and validation of the free-particle Green's function implementation.

For the evaluation of free-particle Green's function matrix elements over PW-SGTOs, the same asymptotic treatments discussed for standard SGTOs must also be applied. However, additional numerical complications may arise from the \( \mathbf{k} \)-dependent normalization factors
\(
\mathcal{N}_{\mathbf{k}_1}(\alpha)\) and \(
\mathcal{N}_{\mathbf{k}_2}(\beta)
\), present in Eqs.~\eqref{eq77} and~\eqref{eq90}.
For small Gaussian exponents, the imaginary part of the complex displacement vector \(\mathbf{R}^{\dagger}\) becomes large, while the normalization factors \( \mathcal{N}_{\mathbf{k}} \) become exponentially small. In such cases, it is essential to include the exponential contribution from the normalization factors directly into the evaluation of the radial integrals in order to avoid numerical instabilities caused by overflow in independent evaluation of exponentially large and small terms.
A similar issue may also appear for the kinetic-energy operator and in that case the problem can also be avoided straightforwardly by incorporating the normalization factors directly into the radial integral \(I_l\).

\section{\label{sec:level-VIII}Numerical illustration}

\subsection{Implementation and validation}

\begin{table}
\caption{\label{Table1} Parameters of the real spherical Gaussian basis functions used to generate reference matrix elements of the free-particle Green's function$^{a,b,c}$.}
\begin{ruledtabular}
\begin{tabular}{ccccc}
 & \multicolumn{2}{c}{Set A} & \multicolumn{2}{c}{Set B} \\
 \hline
$(l,\mu)$ & Number & Exponent & Number & Exponent \\
\hline
$(0, 0)$ & 1$A$ & 5.0 & 1$B$ & 4.5 \\
$(1, 1)$ & 2$A$ & 3.0 & 2$B$ & 2.5 \\
$(2,-2)$ & 3$A$ & 2.0 & 3$B$ & 1.5 \\
$(2, 0)$ & 4$A$ & 2.0 & 4$B$ & 1.5 \\
$(3,-2)$ & 5$A$ & 1.0 & 5$B$ & 0.5 \\
$(3, 0)$ & 6$A$ & 1.0 & 6$B$ & 0.5 \\
$(3, 3)$ & 7$A$ & 1.0 & 7$B$ & 0.5 \\
\end{tabular}
\end{ruledtabular}
\begin{flushleft}
\footnotesize
$^{a}$ Spherical Gaussians are centered at $(-0.1,-0.3,-0.5)$ a.u. and $(1.0,1.6,2.2)$ a.u. for set $A$ and $B$, respectively.\\
$^{b}$ $k_0=0.85215$ a.u.\\
$^{c}$ for PW-SGTOs, the vectors \(\mathbf{k}_1\) and \(\mathbf{k}_2\) are given by
\((0.25,0.50,0.75)\) a.u. and \((0.15,0.30,0.45)\) a.u., respectively.
\end{flushleft}
\end{table}

\begin{table*}
\caption{\label{Table2}
Numerical values of the one-center free-particle Green's function matrix elements
\(G_{l_a\mu_a}^{l_b\mu_b}(\alpha,\beta,k_0)\)
over real SGTOs. The first value corresponds to the real part of \(G_{l_a\mu_a}^{l_b\mu_b}\), while the second value corresponds to the imaginary part. All off-diagonal matrix elements vanish in the one-center case. Numbers in parentheses denote powers of 10.
}
\small
\begin{ruledtabular}
\setlength{\tabcolsep}{6pt}
\begin{tabular}{lcclcc}
Gaussians & \multicolumn{2}{c}{Matrix element} & Gaussians & \multicolumn{2}{c}{Matrix element} \\
\hline
1$A$-1$A$  & $-0.17231976(+00)$  & $-0.88835033(-01)$  & 5$A$-5$A$  & $-0.17018199(+00)$  & $-0.27089402(-02)$ \\
2$A$-2$A$  & $-0.13199662(+00)$  & $-0.14693407(-01)$  & 6$A$-6$A$  & $-0.17018199(+00)$  & $-0.27089402(-02)$  \\
3$A$-3$A$  & $-0.11508121(+00)$  & $-0.27675926(-02)$  & 7$A$-7$A$  & $-0.17018199(+00)$  & $-0.27089402(-02)$  \\
4$A$-4$A$  & $-0.11508121(+00)$  & $-0.27675926(-02)$  &            &                 &                \\
\end{tabular}
\end{ruledtabular}
\end{table*}

\begin{table*}
\caption{\label{Table3}Same as Table~\ref{Table2}, but for two-center free-particle Green's function matrix elements \(G_{l_a\mu_a}^{l_b\mu_b}(\alpha,\beta,\mathbf{R}, k_0)\).}
\small
\begin{ruledtabular}
\setlength{\tabcolsep}{6pt}
\begin{tabular}{lrr lrr}
Gaussians & \multicolumn{2}{c}{Matrix element} & Gaussians & \multicolumn{2}{c}{Matrix element} \\
\hline
1$A$-1$B$  & $ 0.31789853(-01)$ & $-0.56587447(-02)$  & 3$A$-4$B$  & $-0.48695279(-02)$ & $-0.81087363(-03)$ \\
1$A$-2$B$  & $-0.12974029(-02)$ & $0.86379362(-02)$   & 3$A$-5$B$  & $-0.13425038(-02)$ & $ 0.33815825(-02)$ \\
1$A$-3$B$  & $-0.46854317(-02)$ & $-0.49951995(-02)$  & 3$A$-6$B$  & 
$ 0.21531049(-01)$ & $ 0.16340095(-02)$ \\
1$A$-4$B$  & $-0.63162967(-02)$ & $-0.67338858(-02)$  & 3$A$-7$B$  & $-0.64201700(-02)$ & $-0.38874576(-02)$ \\
1$A$-5$B$  & $ 0.40978810(-01)$ & $ 0.12558920(-01)$   & 4$A$-4$B$ & 
$ 0.54344659(-02)$ & $ 0.68052688(-03)$ \\
1$A$-6$B$  & $ 0.30375168(-03)$ & $ 0.93091845(-04)$   & 4$A$-5$B$ & 
$ 0.17372481(-01)$ & $ 0.20955722(-03)$ \\
1$A$-7$B$  & $-0.15685972(-01)$  & $-0.48073350(-02)$  & 4$A$-6$B$  & $-0.19317282(-01)$ & $ 0.43028008(-02)$ \\
2$A$-2$B$  & $ 0.24971701(-02)$  & $-0.48684907(-02)$  & 4$A$-7$B$  & $-0.15823512(-01)$ & $-0.77901870(-03)$ \\
2$A$-3$B$  & $ 0.15645602(-02)$  & $ 0.31967146(-02)$  & 5$A$-5$B$  & $-0.38666440(-02)$ & $-0.10116583(-02)$ \\
2$A$-4$B$  & $-0.38139327(-02)$  & $-0.20057585(-02)$  & 5$A$-6$B$  & 
$ 0.18697634(-01)$ & $-0.12213166(-02)$ \\
2$A$-5$B$  & $-0.16132386(-01)$  & $-0.84197880(-02)$  & 5$A$-7$B$  & $-0.15968986(-01)$ & $-0.30206359(-02)$ \\
2$A$-6$B$  & $ 0.13666201(-01)$  & $ 0.42399592(-02)$   & 6$A$-6$B$  & $-0.47712178(-02)$ & $ 0.13200709(-02)$ \\
2$A$-7$B$  & $ 0.32632091(-02)$  & $ 0.23141531(-02)$   & 6$A$-7$B$  & $-0.33699240(-01)$ & $-0.54891791(-03)$ \\
3$A$-3$B$  & $ 0.67466994(-03)$  & $-0.10272784(-02)$  & 7$A$-7$B$  & 
$ 0.10963089(-01)$ & $-0.26292363(-02)$ \\
\end{tabular}
\end{ruledtabular}
\end{table*}

\begin{table*}
\caption{\label{Table4} Same as Table~\ref{Table2}, but for free-particle Green's function matrix elements \(G_{l_a\mu_a}^{l_b\mu_b}(\alpha,\beta, \mathbf{R}^{\dagger},\mathbf{k}_1,\mathbf{k}_2,k_0)\) evaluated over PW-SGTO basis functions.}
\small
\begin{ruledtabular}
\setlength{\tabcolsep}{6pt}
\begin{tabular}{lrr lrr}
Gaussians & \multicolumn{2}{c}{Matrix element} & Gaussians & \multicolumn{2}{c}{Matrix element} \\
\hline
1$A$-1$B$  & $ 0.25926478(-01)$  & $-0.58213104(-02)$  & 3$A$-4$B$  & $-0.34925853(-03)$  & $ 0.85023126(-02)$ \\
1$A$-2$B$  & $ 0.32321687(-03)$  & $ 0.10368604(-01)$  & 3$A$-5$B$  & $-0.27348113(-02)$  & $ 0.23781267(-01)$ \\
1$A$-3$B$  & $-0.10695467(-01)$  & $-0.32217406(-02)$  & 3$A$-6$B$  & $-0.50841107(-02)$  & $-0.30458293(-01)$ \\
1$A$-4$B$  & $-0.16328752(-01)$  & $-0.44755692(-02)$  & 3$A$-7$B$  & $-0.37894749(-02)$  & $-0.61844550(-02)$ \\
1$A$-5$B$  & $ 0.44504167(-01)$  & $-0.52983416(-01)$  & 4$A$-4$B$  & 
$ 0.150558210(-01)$  & $-0.45011624(-02)$ \\
1$A$-6$B$  & $ 0.20848583(-02)$  & $-0.36174488(-02)$  & 4$A$-5$B$  & $-0.30546150(-01)$  & $-0.15285939(-01)$ \\
1$A$-7$B$  & $-0.16949783(-01)$  & $ 0.20124301(-01)$  & 4$A$-6$B$  & 
$ 0.22027402(-01)$  & $ 0.52001326(-01)$ \\
2$A$-2$B$  & $ 0.40191163(-02)$  & $-0.37752587(-02)$  & 4$A$-7$B$  & 
$ 0.13386420(-01)$  & $ 0.19063587(-01)$ \\
2$A$-3$B$  & $ 0.20814693(-02)$  & $ 0.52507741(-02)$  & 5$A$-5$B$  & 
$ 0.51590889(-01)$  & $ 0.23313885(-01)$ \\
2$A$-4$B$  & $-0.58094712(-02)$  & $ 0.46428166(-02)$  & 5$A$-6$B$  & 
$-0.44819524(-01)$  & $-0.13229320(-01)$ \\
2$A$-5$B$  & $-0.30218867(-01)$  & $-0.22046960(-03)$  & 5$A$-7$B$  & 
$-0.19945042(-01)$  & $ 0.56960950(-02)$ \\
2$A$-6$B$  & $ 0.95876909(-02)$  & $-0.12401086(-01)$  & 6$A$-6$B$  & 
$ 0.91631553(-01)$  & $-0.18106976(-01)$ \\
2$A$-7$B$  & $ 0.93272016(-02)$  & $ 0.23520845(-02)$  & 6$A$-7$B$  & 
$ 0.40542996(-01)$  & $ 0.30205007(-01)$ \\
3$A$-3$B$  & $ 0.46662722(-02)$  & $-0.17130418(-02)$  & 7$A$-7$B$  & 
$ 0.56578290(-02)$  & $-0.18534249(-01)$ \\
\end{tabular}
\end{ruledtabular}
\end{table*}
The pilot implementation of the free-particle Green's function matrix elements over SGTO and PW-SGTO basis functions with real spherical harmonics was first carried out in \textsc{Mathematica}\cite{Mathematica13.3}. This provided high-precision reference values for validating the numerical implementation. The production implementation was subsequently done within the \textsc{Libqints} integral library which is a part of the \textsc{Q-Chem} quantum chemistry package.

The correctness of the implementation was verified and cross-checked in several independent ways. First, the matrix elements computed in \textsc{Libqints} were compared directly against the high-precision reference values obtained from \textsc{Mathematica}. In addition, an analogous implementation was developed for kinetic-energy matrix elements. These integrals were independently verified against the standard recurrence relations already available in \textsc{Libqints} for real spherical harmonic Gaussian basis functions. This provided an important consistency check for the angular part of the formalism, since both the kinetic-energy and Green's function operators share the same angular-coupling structure, differing only in the form of the radial momentum integral \(I_l(\eta,R)\).

Finally, the computed matrix elements \(G_{l_a\mu_a}^{l_b\mu_b}(\alpha,\beta, \mathbf{R}, k_0)\) were benchmarked against the reference data reported by Čársky \textit{et al.}~\cite{Carsky1996}. Since the results of Ref.~\cite{Carsky1996} were given in the Cartesian Gaussian basis, the corresponding quantities were transformed into the spherical harmonic representation using Cartesian-to-spherical transformations\cite{Schlegel1995, ribaldone2024}. The successful agreement obtained in all these tests provides strong confidence in both the correctness of the derived formulas and the numerical implementation.

To facilitate future implementations and independent verification, we also provide a set of numerical benchmark values for both one-center and two-center matrix elements. Table~\ref{Table1} summarizes the Gaussian parameters used to generate the benchmark data. Tables~\ref{Table2} and \ref{Table3} contain reference values for one-center $G_{l_a\mu_a}^{l_b\mu_b}(\alpha,\beta, k_0)$ and two-center \(G_{l_a\mu_a}^{l_b\mu_b}(\alpha,\beta, \mathbf{R}, k_0)\) matrix elements over real SGTO basis functions, respectively. Finally, Table~\ref{Table4} presents reference values for matrix elements \(G_{l_a\mu_a}^{l_b\mu_b}(\alpha,\beta, \mathbf{R}^{\dagger},\mathbf{k}_1,\mathbf{k}_2,k_0)\) evaluated in the PW-SGTO basis.

\subsection{Variation and asymptotic behavior of the matrix elements}
\begin{figure*}[t]
\centering
\includegraphics[width=\textwidth]{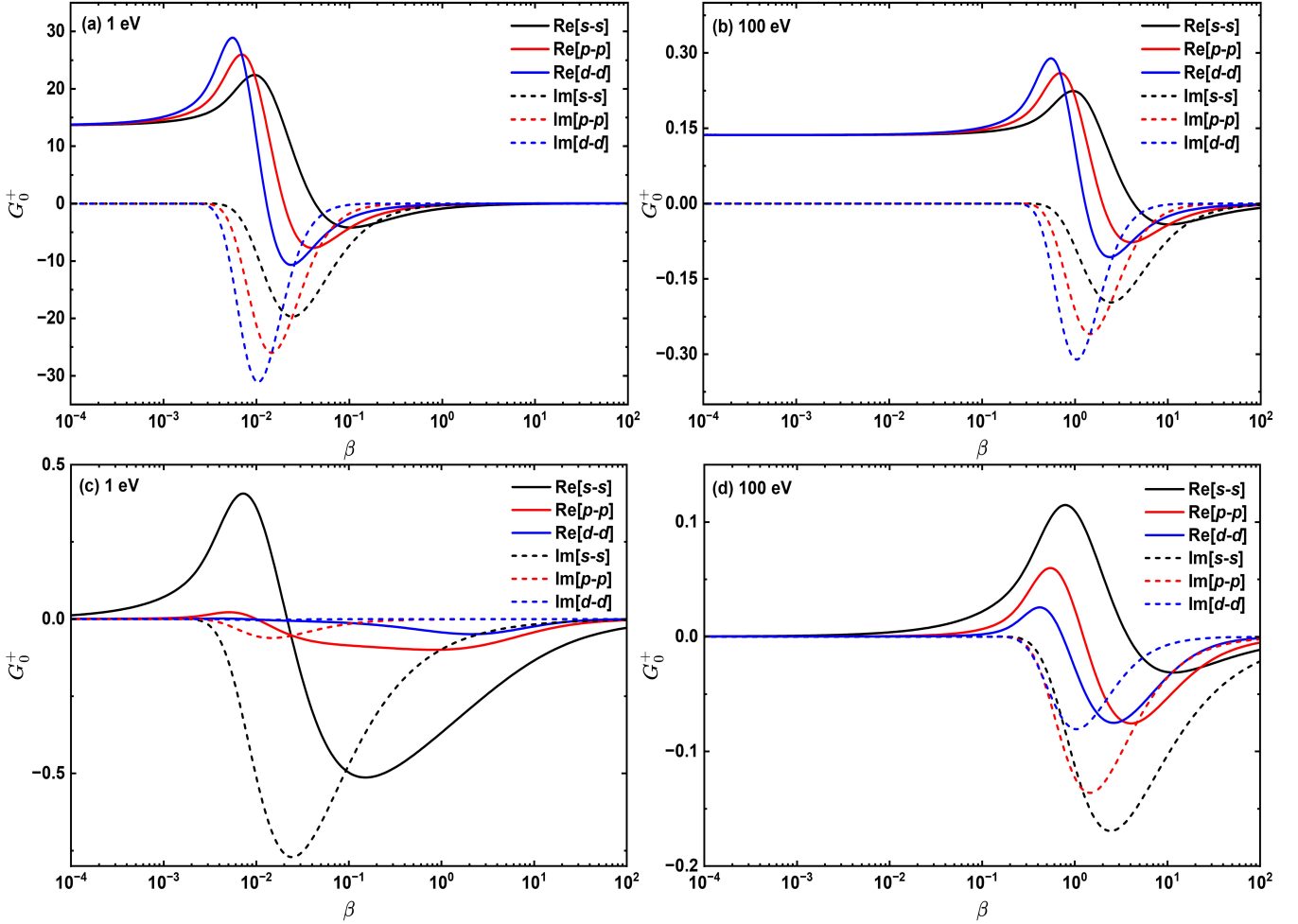}
\caption{\label{fig:3} One-center free-particle Green's function matrix elements
\(G_{l\mu}^{l\mu}\)
for \(s\)-, \(p\)-, and \(d\)-type spherical Gaussian functions corresponding to electron energies of  1  and 100 eV. The upper panel shows diagonal matrix elements (\(\alpha=\beta\)), while the lower panel presents off-diagonal matrix elements with \(\alpha\) fixed at 5.0.}
\end{figure*}

\begin{figure*}[t]
\centering
\includegraphics[width=16.2cm,height=6.2cm]{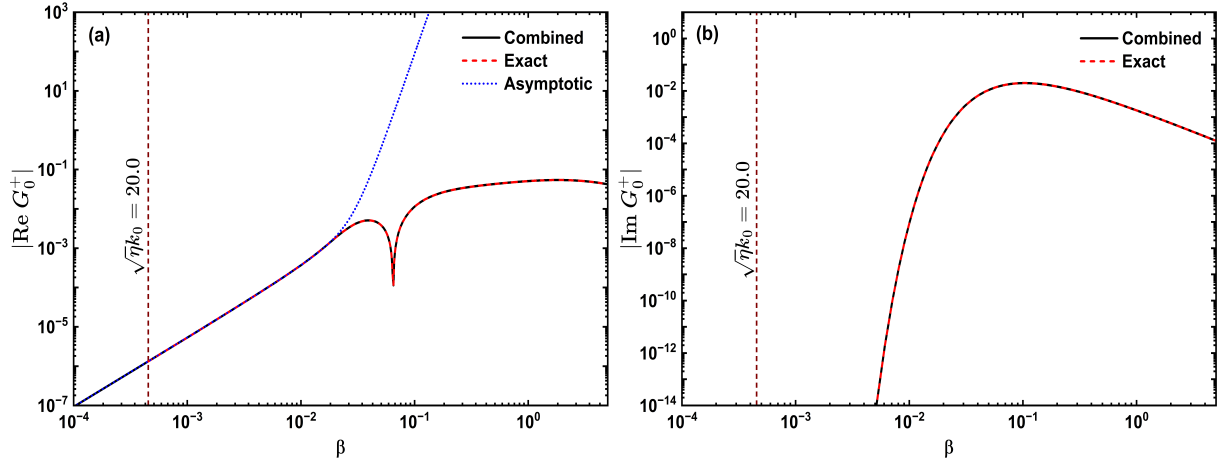}
\caption{\label{fig:1}
One-center free-particle Green's function matrix element \(G_{l_a\mu_a}^{l_b\mu_b}\) with
\((l_a,\mu_a)=(l_b,\mu_b)=(2,0)\)
as a function of the Gaussian exponent \(\beta\), with the first exponent fixed at \(\alpha=5.0\).
Panel (a) shows the absolute value of the real part, while panel (b) shows the absolute value of the imaginary part.
In panel (a), the blue line corresponds to the asymptotic expression given by Eq.~\eqref{eq101}. The black solid line shows the full numerical result obtained using the switching condition \(\sqrt{\eta}\,k_0=20\) for the asymptotic treatment.}
\end{figure*}

\begin{figure*}[t]
\centering
\includegraphics[width=16.2cm,height=6.2cm]{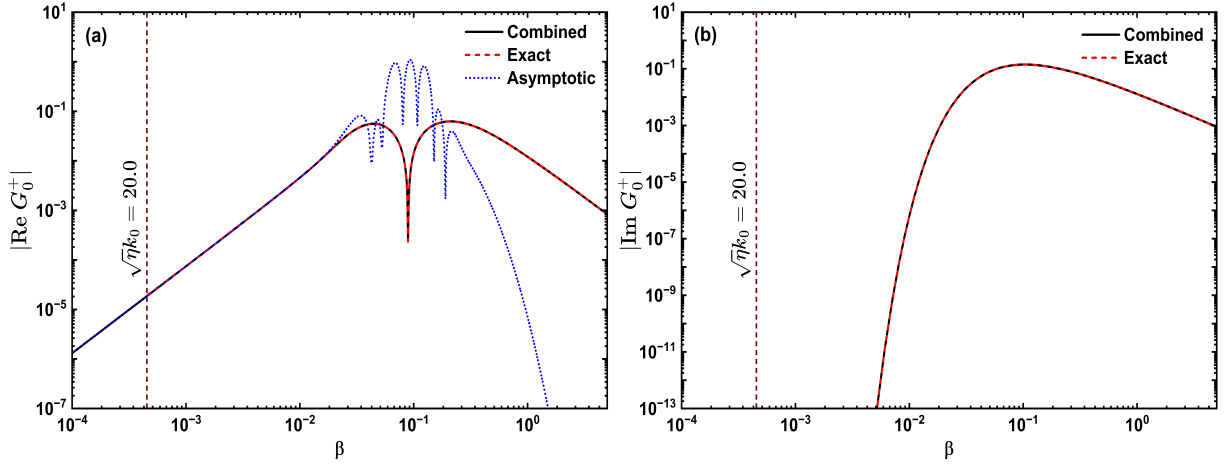}
\caption{\label{fig:2}Two-center free-particle Green's function matrix element
\(G_{l_a\mu_a}^{l_b\mu_b}\)
with \((l_a,\mu_a)=(0,0)\) and \((l_b,\mu_b)=(2,0)\),
shown as a function of the Gaussian exponent \(\beta\), while the first exponent is fixed at \(\alpha=5.0\). The Gaussian centers are as those given in Table~\ref{Table1}.
Panel~(a) shows the absolute value of the real part, whereas panel~(b) shows the absolute value of the imaginary part. In panel~(a), the blue line corresponds to the asymptotic expression of Eq.~\eqref{eqd19}, while the black solid line represents the full numerical result obtained using the switching condition \(\sqrt{\eta}\,k_0=20\) for the asymptotic treatment.}
\end{figure*}

For illustrative purposes, we analyze the magnitude and qualitative behavior of the free-particle Green's function matrix elements $G_{l_a\mu_a}^{l_b\mu_b}$ for different electron energies and Gaussian exponents. A representative example for the one-center diagonal (\(\alpha=\beta\)) and off-diagonal matrix elements  is shown in Fig.~\ref{fig:3}, where the real and imaginary parts of $G_{l_a\mu_a}^{l_b\mu_b}$  are plotted for two electron energies of 1 and  100 eV.

The dependence on the Gaussian exponent reflects how efficiently a Gaussian function of a given spatial extent couples to the continuum at a particular electron energy. 
In the case of diagonal matrix elements, the real part approaches a constant value for very diffuse Gaussian functions (proportional to \(1/k_0^2\)), while it vanishes for very tight Gaussians. In contrast, the imaginary part tends to zero in both the diffuse and tight Gaussian limits, exhibiting a pronounced maximum when the Gaussian width becomes comparable to the electron de Broglie wavelength. 
The off-diagonal matrix elements are generally smaller in magnitude than the corresponding diagonal elements and approach zero in both the diffuse and tight Gaussian limits.

The imaginary part displays a negative peak originating from on-shell continuum contributions, whereas the real part exhibits a dispersive-like behavior accompanied by a characteristic sign change. As the electron energy increases, the maxima shift toward larger values of exponent \(\beta\), indicating that higher-energy continuum electrons require tighter Gaussian functions for efficient representation. 

The trends from Fig.~\ref{fig:3} also provide useful guidance for constructing optimized Gaussian basis sets for representing the Green's  operator using SGTO expansions. In particular, even-tempered exponent distributions appear to be especially suitable, provided that a sufficient density of exponents is included in the region where the Green's function matrix elements exhibit the largest variation and magnitude. Consequently, analysis of the matrix-element profiles can help identify apriori the most efficient Gaussian exponents to represent the Green's function  at a given electron energy.

As discussed at the end of Section~\ref{sec:level-II}, practical applications of SGTO and PW-SGTO basis sets for representing the free-particle Green's function require the use of asymptotic formulas in the regime
$\sqrt{\eta}\,k_0 \gg 1$ in order to avoid numerical instabilities associated with the direct evaluation of the complementary error function. In this regime, straightforward numerical evaluation may lead to overflow errors or undefined numerical values due to the presence of exponentially large and small contributions.

The asymptotic behavior is illustrated in Fig.~\ref{fig:1} for one-center matrix elements and in Fig.~\ref{fig:2} for the two-center case. In both figures, the asymptotic expressions of Eqs.~\eqref{eq101} and~\eqref{eqd19} accurately reproduce the exact matrix elements in the diffuse-Gaussian regime corresponding to small Gaussian exponents. In the present implementation, the transition from the exact expressions to the asymptotic formulas is performed at the switching condition$\sqrt{\eta}\,k_0 = 20$
which represents a rather conservative threshold. The asymptotic expansion was evaluated using terms up to \(n=10\). In both the one- and two-center cases, the imaginary part vanishes once the asymptotic expressions are applied, consequently, no blue asymptotic curve appears in the right panels of Figs.~\ref{fig:1} and~\ref{fig:2}.

\section{\label{sec:level-IX}Conclusions}

An efficient description of electronic continuum states remains one of the major challenges in modern electronic structure theory. This problem is particularly important in the context of rapidly developing novel light sources capable of probing highly excited or ionized states of atoms and molecules, which subsequently decay through autoionization or related continuum-driven processes. Despite the tremendous success of standard quantum chemistry methods for bound electronic states, their application to continuum electrons remains limited. 

In our view, one of the main obstacles preventing broader application of Gaussian-based quantum chemistry methods to continuum problems has been the lack of efficient and general formulations for free-particle Green's function matrix elements in Gaussian basis sets. Previous formulations were often algebraically cumbersome and have therefore seen only very limited practical use.

In the present work, we derived efficient analytical formulas for evaluating matrix elements of the free-particle Green's function  in the basis of spherical Gaussian-type orbitals (SGTOs) and plane-wave-modified spherical Gaussians (PW-SGTOs). The resulting expressions are compact, computationally efficient, and fully general with respect to angular momentum. In addition to the stable recurrence relations developed, we have also analyzed the asymptotic treatments required for robust numerical evaluation over a broad range of Gaussian exponents and electron energies.

We expect that the present formalism will enable direct incorporating continuum-electron description into electronic structure methods that are designed primarily for bound-domain states. In particular, the developed methodology will facilitate solving the Lippmann--Schwinger equation for continuum electron within Gaussian basis representation. Such an approach will naturally incorporates the correct boundary conditions in the interaction region, while providing a  multicenter  description of the continuum orbital, analogous to the treatment commonly used for bound electronic states.

More broadly, the present developments may help bridge the gap between highly accurate bound-state quantum chemistry methods and explicit continuum-electron treatments while retaining the computational advantages of Gaussian basis techniques. This could enable more accurate calculations of processes involving bound--continuum couplings within a single electronic structure software. Specific applications of the present formalism to ab initio modeling of electron--molecule scattering cross sections and various autoionization processes will be reported in forthcoming publications.

\begin{acknowledgments}
We acknowledge  financial support from the Polish National Agency for Academic Exchange through Polish Returns Programme (awarded to W.S.) and  through Ulam fellowship (awarded to D.M.).
\end{acknowledgments}

\appendix
\onecolumngrid
\section{Gaunt coefficients for complex-valued and real spherical harmonics}\label{appG}
The general expression for the Gaunt coefficients can be written in terms of Wigner's 3j symbols as\cite{edmonds1996}:
\begin{eqnarray}\label{eq22}
\langle lm | l_a m_a | l_b m_b \rangle &=&\int d\hat{\mathbf{r}} Y^*_{lm}(\hat{\mathbf{r}})Y_{l_am_a}(\hat{\mathbf{r}})Y_{l_bm_b}(\hat{\mathbf{r}})\nonumber \\
&=&(-1)^m \sqrt{\frac{[l_a][l_b][l]}{4\pi}}
\begin{pmatrix}
l & l_a & l_b \\
0 & 0 & 0
\end{pmatrix}
\begin{pmatrix}
l & l_a & l_b \\
-m & m_a & m_b
\end{pmatrix}.
\end{eqnarray}
where, $[l_{e}]$ is defined as $2l_e+1$, with $l_e=l,l_a$ and $l_b$.  
In our derivation, Gaunt coefficients arise naturally from the product of solid harmonics appearing in Eq.~\eqref{eq12}:
\begin{equation}\label{eq16}
  \mathcal{Y}^*_{l_am_a}(\mathbf{q})\mathcal{Y}_{l_bm_b}(\mathbf{q})=\sum_{lm}\langle l_bm_b|l_am_a|lm \rangle q^{2n}\mathcal{Y}_{lm}(\mathbf{q}),
\end{equation}
Here $n=(l_a+l_b-l)/2$ is a non-negative integer value. According to the selection rule of the Gaunt coefficients, the allowed $l$ values are those satisfying  $l_a+l_b+l=$ even integer and the triangular rule $\Delta(l_al_bl)$, i.e.
    \begin{equation}\label{eq18}
         l\in [l_{max},l_{max}-2,....,l_{min}],
    \end{equation}
with
    \begin{equation}\label{eq19}
        l_{max}=l_a+l_b,
    \end{equation}
 and
   \begin{equation}\label{eq20}
    l_{min} = 
    \begin{cases} 
    \kappa(l_a,l_b,m_a,m_b) & \text{if }\kappa(l_a,l_b,m_a,m_b)+l_{max} \text{ is even }\\
     \kappa(l_a,l_b,m_a,m_b)+1 & \text{if }\kappa(l_a,l_b,m_a,m_b)+l_{max} \text{ is odd }
    \end{cases}
    \end{equation}
where 
    \begin{equation}\label{eq21}
         \kappa(l_a,l_b,m_a,m_b)=\text{max}(|l_a-l_b|,|m_a+m_b|).
    \end{equation}

In the case of real spherical harmonics, the corresponding Gaunt coefficients $\langle l\mu|l_a\mu_a|l_b\mu_b \rangle$ were analyzed by Homeier and Steinborn \cite{HOMEIER1996}. These coefficients involve three possible combinations of angular momentum quantum numbers:
  \begin{enumerate}
    \item $\mu$ and $\mu_a$ and $\mu_b$ are non zero, 
    \begin{eqnarray}\label{eq:29}
    \langle l\mu|l_a\mu_a|l_b\mu_b \rangle&=& 2\langle l\mu_a+\mu_b|l_a\mu_a|l_b\mu_b \rangle\Re\Big([U^{\mu}_{l\mu_a+\mu_b}]^* U^{\mu_a}_{l_a\mu_a}U^{\mu_b}_{l_b\mu_b}\Big)\nonumber \\
    &&+2\langle l\mu_a-\mu_b|l_a\mu_a|l_b-\mu_{b} \rangle\Re\Big([U^{\mu}_{l\mu_a-\mu_b}]^* U^{\mu_a}_{l_a\mu_a}U^{\mu_b}_{l_b-\mu_b}\Big),
    \end{eqnarray}
    \item  one of $\mu_a$ or $\mu_b$ is zero \\
    (a) if $\mu\neq0$ and $\mu_a\neq0$ and $\mu_b=0$
    \begin{equation}\label{eq:30}
        \langle l\mu|l_a\mu_a|l_b\mu_b \rangle= 2\langle l\mu_a|l_a\mu_a|l_b0 \rangle\Re\Big([U^{\mu}_{l\mu_a}]^* U^{\mu_a}_{l_a\mu_a}\Big),
\end{equation}
  (b) if $\mu\neq0$ and $\mu_a=0$ and $\mu_b\neq0$
    \begin{equation}\label{eq:31}
        \langle l\mu|l_a\mu_a|l_b\mu_b \rangle= 2\langle l\mu_a|l_a0|l_b\mu_b \rangle\Re\Big([U^{\mu}_{l\mu_b}]^* U^{\mu_b}_{l_b\mu_b}\Big).
\end{equation}
\item if $\mu_a=\mu_b=0$ 
    \begin{equation}\label{eq:32}
    \langle l\mu|l_a 0|l_b 0 \rangle= \delta_{\mu 0}\langle l 0|l_a 0|l_b 0 \rangle.
\end{equation}
  \end{enumerate}
The above expressions involve the unitary transformation matrix \(U^{\mu}_{lm}\), defined as
\begin{eqnarray}\label{eq33}
    U^{\mu}_{lm}&=&\delta_{m0}\delta_{\mu0}+\frac{1}{\sqrt{2}}\Big(\Theta(\mu)\delta_{m\mu}+\Theta(-\mu)(+i)(-1)^m\delta_{m\mu}\nonumber\\
        &&+\Theta(-\mu)(-i)\delta_{m-\mu}+\Theta(\mu)(-1)^m\delta_{m-\mu}\Big),
\end{eqnarray}
where \(\delta_{m\mu}\) denotes the Kronecker delta and
\begin{equation}\label{eq34}
\Theta(\mu) = 
\begin{cases} 
1 & \text{for } \mu > 0, \\
0 & \text{for } \mu \leq 0.
\end{cases}
\end{equation}
The definition in Eq.~\eqref{eq34} ensures that \(U^{\mu}_{lm}=0\) for \(|\mu|\neq|m|\). The selection rules for \(l\) remain identical to those given in Eqs.~\eqref{eq18}--\eqref{eq21}, while \(\mu\in[-l,\ldots,l]\). However, only the combinations
\[
\mu=\{\mu_a+\mu_b,\mu_a-\mu_b,-\mu_a+\mu_b,-\mu_a-\mu_b\}
\]
give nonvanishing contributions to the real Gaunt coefficients.

\section{\label{appK}Matrix elements of kinetic energy operator}
Here we provide recurrence relations for the radial momentum integrals appearing in kinetic-energy matrix elements. These relations are closely analogous to those derived in Sec.~2 for the free-particle Green's function. 

The general expression for the kinetic-energy matrix elements in spherical Gaussian-type orbitals with complex spherical harmonics, based on Eq.~\eqref{eq25}, is
\begin{equation}\label{eq37}
T_{l_a m_a}^{l_b m_b}(\alpha,\beta,\mathbf{R}) = 4\pi C_{l_al_b}(\alpha,\beta) \sum_{lm} i^l \langle lm|l_a m_a|l_b m_b\rangle Y_{lm}(\widehat{\mathbf{R}}) I_l^P(a,R).
\end{equation}
or with real spherical harmonics, by applying Eq.~\eqref{eq27}:
\begin{equation}\label{eq39}
T_{l_a\mu_a}^{l_b\mu_b}(\alpha,\beta,\mathbf{R}) = 4\pi C_{l_al_b}(\alpha,\beta) \sum_{l\mu} i^l \langle l\mu|l_a\mu_a|l_b\mu_b\rangle X_{l}^{\mu}(\widehat{\mathbf{R}}) I_l^P(\eta,R).
\end{equation}
Here the radial momentum integral is defined as
\begin{equation}\label{eq38}
I_l^P(\eta,R) = \frac{1}{2} \int_{0}^{\infty} q^{P}e^{-\eta q^2}j_l(qR)\,dq,
\end{equation}
with $P=4+l_a+l_b$.

We again employ the same recurrence relation for spherical Bessel functions as given in Eq.~\eqref{eq40},
\begin{equation}\label{eq40b}
I_{l+1}^{P+1}(\eta,R) = \frac{2l+1}{R}I_l^P(\eta,R) - I_{l-1}^{P+1}(\eta,R), \qquad l\ge1.
\end{equation}

As before, it is sufficient to evaluate only the integrals corresponding to \(l=0\) and \(l=1\), while all higher-order terms can be generated recursively from Eq.~\eqref{eq40b}. The required radial integrals \(I_l^P(\eta,R)\) may be obtained through differentiation of lower-order integrals with respect to \(\eta\) and \(R\). In practice, the evaluation becomes particularly efficient once the fundamental integrals \(I_0^4\) and \(I_1^5\) are known. Higher-order terms can then be generated solely through differentiation with respect to \(\eta\). The general recurrence formula for the kinetic-energy integrals is therefore
\begin{equation}\label{eq41}
I_l^{4+2k+l}(\eta,R) = (-1)^k \sum_{q=0}^k \binom{k}{q} f_l^{(k-q)}(\eta,R) g_l^{(q)}(\eta,R),
\end{equation}
where \(k\in\{0,1,2,\ldots\}\).

Equation~\eqref{eq41} generates all integrals of the form
\(I_0^4\), \(I_0^6\), \(I_0^8\), \(I_0^{10}\), \(\ldots\) for \(l=0\), and
\(I_1^5\), \(I_1^7\), \(I_1^9\), \(I_1^{11}\), \(\ldots\) for \(l=1\). The remaining higher-order terms, such as \(I_2^6\), \(I_2^8\), \(I_3^7\), \(I_4^8\), etc., can then be obtained from Eq.~\eqref{eq40b}.

The initial terms entering Eq.~\eqref{eq41} are
\begin{equation}\label{eq42} f^{(0)}_0(\eta,R) = \frac{e^{-R^2/4\eta}\sqrt{\pi}}{32\eta^{7/2}},
\end{equation}
\begin{equation}\label{eq43}
g^{(0)}_0(\eta,R) = 6\eta-R^2,
\end{equation}
and
\begin{equation}\label{eq44}
f^{(0)}_1(\eta,R) = \frac{e^{-R^2/4\eta}\sqrt{\pi}}{64\eta^{9/2}},
\end{equation}
\begin{equation}\label{eq45}
g^{(0)}_1(\eta,R) = 10\eta R-R^3.
\end{equation}
Furthermore,
\begin{equation}
g^{(1)}_0(\eta,R)=6,
\qquad
g^{(1)}_1(\eta,R)=10R,
\end{equation}
while all higher-order terms vanish. The higher-order contributions \(f_l^{(n)}(\eta,R)\) for \(n\ge1\) can be generated recursively as
\begin{equation}\label{eq46}
f_l^{(n)}(\eta,R) = -\sum_{q=0}^{n-1} \binom{n-1}{q} f_l^{(n-q-1)}(\eta,R) h_l^{(q)}(\eta,R),
\end{equation}
where
\begin{equation}\label{eq47}
h_l^{(n)}(\eta,R) = (-1)^n \Bigg[ \frac{(7+2l)}{2}n! \, \eta^{-(n+1)} - \frac{R^2}{4}(n+1)! \, \eta^{-(n+2)} \Bigg].
\end{equation}

For one-center kinetic-energy matrix elements (\(R=0\)), the radial integral can be evaluated simply in closed form yielding:
\begin{equation}\label{eq48}
T_{l_am_a}^{l_bm_b}(\alpha,\beta) = \frac{1}{4} C_{l_al_b}(\alpha,\beta) \eta^{\frac{1}{2}(-5-l_a-l_b)}\Gamma\Big[\frac{1}{2}(5+l_a+l_b)\Big]
\delta_{l_al_b}\,
\delta_{m_am_b}.
\end{equation}

\section{Recurrence relations for solid spherical harmonics}\label{appB}
The general formulas for multi-center integrals over spherical Gaussian basis functions depend on solid spherical harmonics and therefore require efficient evaluation of these functions for both real and complex vector arguments. In the case of standard SGTOs, only real vectors appear, whereas PW-SGTOs naturally introduce complex displacement vectors. Here we briefly summarize the basic properties of solid spherical harmonics used throughout the evaluation of the matrix elements.

The regular solid harmonic for a real vector \(\mathbf{R}\) is defined as
\begin{equation}\label{eqb1} \mathcal{Y}_{lm}(\mathbf{R}) = R^l Y_{lm}(\widehat{\mathbf{R}}),
\end{equation}
where \(Y_{lm}(\widehat{\mathbf{R}})\) denotes the spherical harmonic and \(R=|\mathbf{R}|\).
For PW-SGTOs complex vectors \(\mathbf{R}=(X,Y,Z)\in\mathbb{C}^3\)
naturally arise, requiring operations with complex arguments (see p.~71 of Ref.~\cite{biedenharn1984}).
Using the recurrence relations of the associated Legendre polynomials, the following simplified expression for solid harmonics for \(m\ge0\) holds\cite{Kuang21997}: 
\begin{equation}\label{eqb2}
c(l+1, m) \mathcal{Y}_{l+1,m}(\mathbf{R}) = Z \mathcal{Y}_{lm}(\mathbf{R}) - c(l, m) (\mathbf{R} \cdot \mathbf{R}) \mathcal{Y}_{l-1,m}(\mathbf{R}),
\end{equation}
with 
\begin{equation}\label{eqb3}
    c(l, m) = \sqrt{\frac{(l-m)(l+m)}{(2l-1)(2l+1)}}.
\end{equation}
The recurrence relation in Eq.~(\ref{eqb2}) is initialized with the following starting values:
\begin{equation}\label{eqb4}
    \mathcal{Y}_{l-1,l}(\mathbf{R})=0,
\end{equation}
and 
\begin{equation}\label{eqb5}
    \mathcal{Y}_{l,l}(\mathbf{R})=s(l)(X+iY)^l,
\end{equation}
where
\begin{equation}\label{eqb6}
    s(l)=\sqrt{\frac{(-1)^l}{ 2^ll!}\frac{(2l+1)!}{4\pi}}.
\end{equation}
Using the relation
\([\mathcal{Y}_{lm}(\mathbf{R})]^* = (-1)^m \mathcal{Y}_{l,-m}(\mathbf{R}^{*}),\)
the solid spherical harmonics for \(m<0\) can be obtained as
\begin{equation}\label{eqb7}
     \mathcal{Y}_{l,-m}(\mathbf{R})=(-1)^m\Big[\mathcal{Y}_{lm}(\mathbf{R}^{*})\Big]^*.
\end{equation}
Recurrence formulas from above are valid for both real and complex vectors, allowing efficient evaluation of the solid spherical harmonics defined in Eq.~(\ref{eqb1}). The corresponding real solid spherical harmonics can then be obtained directly using Eq.~(\ref{eq28}).

\section{Addition theorem for solid spherical harmonics}\label{appA}
\subsection{Complex solid spherical harmonics}
Addition theorem for the solid spherical harmonics \cite{steinborn1973, homeier1991}:
\begin{equation}\label{eq73}
            \mathcal{Y}_{lm}(\mathbf{r}_1+\mathbf{r_2})=4\pi \sum_{l^{\prime}=0}^l\sum_{m^{\prime}=\text{max}\atop (-l^{\prime},m-l+l^{\prime})}^{\text{min}(l^{\prime},m+l-l^{\prime})}G(lm|l^{\prime}m^{\prime})\mathcal{Y}_{l^{\prime}m^{\prime}}(\mathbf{r_1})\mathcal{Y}_{l-l^{\prime},m-m^{\prime}}(\mathbf{r_2}),
\end{equation}
where $G(lm|l^{\prime}m^{\prime})$ coefficients are defined as:
\begin{equation}\label{eq74}
G(lm|l^{\prime}m^{\prime})=\Bigg[\frac{2l+1}{4\pi(2l^{\prime}+1)[2(l-l^{\prime})+1]}\binom{l+m}{l^{\prime}+m^{\prime}}\binom{l-m}{l^{\prime}-m^{\prime}}\Bigg]^{1/2}.
\end{equation}
\subsection{Real solid spherical harmonics}
 The addition theorem for unnormalized real regular solid harmonics can be in general expressed as \cite{rico2013},
\begin{equation}\label{eq84}
            z_{l}^{\mu}(\mathbf{r}_1+\mathbf{r}_2)=\sum_{l^{\prime}=0}^{l}\sum_{\mu^{\prime}=-l^{\prime}}^{l^{\prime}}z_{l^{\prime}}^{\mu^{\prime}}(\mathbf{r}_1)\sum_{\mu^{\prime\prime}}z_{l-l^{\prime}}^{\mu^{\prime\prime}}(\mathbf{r}_2)c_{l^{\prime} \mu^{\prime} \mu^{\prime\prime}}^{l\mu},
\end{equation}
where the unnormalized real harmonic $z_{l}^{\mu}(\mathbf{r})$ is defined in Eq.~\eqref{eq85} and \eqref{eq86}. To provide explicit expressions for the coefficients \(c_{l^{\prime}\mu^{\prime}\mu^{\prime\prime}}^{l\mu}\), the real spherical harmonics must be treated separately for the following two cases:
\begin{enumerate}
    \item for $\mu\ge0$
    \begin{eqnarray}\label{eqa1}
        z_l^{\mu}(\mathbf{r_1}+\mathbf{r_2}) &=& \sum_{l^{\prime}=0}^{l} \sum_{\mu^{\prime}=\max\atop {(0,l{\prime}-l+\mu)}}^{\min(l^{\prime},\mu)} A_{l^{\prime}\mu^{\prime}}^{l\mu} \Big[ z_{l^{\prime}}^{\mu^{\prime}}(\mathbf{r_1}) z_{l-l^{\prime}}^{\mu-\mu^{\prime}}(\mathbf{r_2})\nonumber \\
        &&\quad - (1-\delta_{\mu^{\prime},0})(1-\delta_{\mu^{\prime},\mu}) z_{l^{\prime}}^{-\mu^{\prime}}(\mathbf{r_1}) z_{l-l^{\prime}}^{-(\mu-\mu^{\prime})}(\mathbf{r_2}) \Big]\nonumber \\
        &&+ \sum_{l^{\prime}=\mu+1}^{l-1} \sum_{\mu^{\prime}=\mu+1}^{\min(l^{\prime},l-^{\prime}+\mu)} B_{l^{\prime}\mu^{\prime}}^{l\mu} \Big[ z_{l^{\prime}}^{\mu^{\prime}}(\mathbf{r_1}) z_{l-l^{\prime}}^{\mu^{\prime}-\mu}(\mathbf{r_2})\nonumber \\
        &&\quad + z_{l^{\prime}}^{-\mu^{\prime}}(\mathbf{r_1}) z_{l-l^{\prime}}^{-(\mu^{\prime}-\mu)}(\mathbf{r_2}) \Big]\nonumber \\
        &&+ \sum_{l^{\prime}=1}^{l-\mu-1} \sum_{\mu^{\prime}=1}^{\min(l^{\prime},l-l^{\prime}-\mu)} C_{l^{\prime}\mu^{\prime}}^{l\mu} \Big[ z_{l^{\prime}}^{\mu^{\prime}}(\mathbf{r_1}) z_{l-l^{\prime}}^{\mu^{\prime}+\mu}(\mathbf{r_2})\nonumber \\
        &&\quad + z_{l^{\prime}}^{-\mu^{\prime}}(\mathbf{r_1}) z_{l-l^{\prime}}^{-(\mu^{\prime}+\mu)}(\mathbf{r_2}) \Big],
    \end{eqnarray}
    \item for $\mu<0$
\begin{eqnarray}\label{eqa2}
     z_l^{-|\mu|}(\mathbf{r_1}+\mathbf{r_2}) &=& \sum_{l^{\prime}=0}^{l} \sum_{\mu^{\prime}=\max\atop {(0,l^{\prime}-l+|\mu|)}}^{\min(l^{\prime},|\mu|)} A_{l^{\prime}\mu^{\prime}}^{l|\mu|} \Big[ (1-\delta_{\mu^{\prime},|\mu|}) z_{l^{\prime}}^{\mu^{\prime}}(\mathbf{r_1}) z_{l-l^{\prime}}^{-(|\mu|-\mu^{\prime})}(\mathbf{r_2})\nonumber \\
    &&\quad + (1-\delta_{\mu^{\prime},0}) z_{l^{\prime}}^{-\mu^{\prime}}(\mathbf{r_1}) z_{l-l^{\prime}}^{(|\mu|-\mu^{\prime})}(\mathbf{r_2}) \Big]\nonumber \\
    &&+ \sum_{l^{\prime}=|\mu|+1}^{l-1} \sum_{\mu^{\prime}=|\mu|+1}^{\min(l^{\prime},l-l^{\prime}+|\mu|)} B_{l^{\prime}\mu^{\prime}}^{l|\mu|} \Big[ -z_{l^{\prime}}^{\mu^{\prime}}(\mathbf{r_1}) z_{l-l^{\prime}}^{-(\mu^{\prime}-|\mu|)}(\mathbf{r_2})\nonumber \\
    &&\quad + z_{l^{\prime}}^{-\mu^{\prime}}(\mathbf{r_1}) z_{l-l^{\prime}}^{(\mu^{\prime}-|\mu|)}(\mathbf{r_2}) \Big]\nonumber \\
    &&+ \sum_{l^{\prime}=1}^{l-|\mu|-1} \sum_{\mu^{\prime}=1}^{\min(l^{\prime},l-l^{\prime}-|\mu|)} C_{l^{\prime}\mu^{\prime}}^{l|\mu|} \Big[ z_{l^{\prime}}^{\mu^{\prime}}(\mathbf{r_1}) z_{l-l^{\prime}}^{-(\mu^{\prime}+|\mu|)}(\mathbf{r_2})\nonumber \\
    &&\quad - z_{l^{\prime}}^{-\mu^{\prime}}(\mathbf{r_1}) z_{l-l^{\prime}}^{(\mu^{\prime}+|\mu|)}(\mathbf{r_2}) \Big],
\end{eqnarray}

where
\begin{equation}\label{eqa3}
A_{l^{\prime}\mu^{\prime}}^{l\mu}=\frac{(l+\mu)!}{(l^{\prime}+\mu^{\prime})!(l+\mu-l^{\prime}-\mu^{\prime})!},
\end{equation}
\begin{equation}\label{eqa4}
B_{l^{\prime}\mu^{\prime}}^{l\mu}=\frac{(-1)^{\mu^{\prime}-\mu}(l+\mu)!}{(l^{\prime}+\mu^{\prime})!(l-\mu-l^{\prime}+\mu^{\prime})!},
\end{equation}
and
\begin{equation}\label{eqa5}
C_{l^{\prime}\mu^{\prime}}^{l\mu}=\frac{(-1)^{\mu^{\prime}}(l+\mu)!}{(l^{\prime}+\mu^{\prime})!(l+\mu-l^{\prime}+\mu^{\prime})!}.
\end{equation}
\end{enumerate}
Using the above addition theorem for real spherical harmonics together with the corresponding definitions, the PW-SGTO product can be expressed as
\begin{enumerate}
    \item for $\mu\ge0$
    \begin{eqnarray}\label{eqa6}
        e^{i\mathbf{k}\cdot\mathbf{r}}\phi_{l}^{\mu}(\alpha,\mathbf{r}-\mathbf{C})&=&N_{l}(\alpha)N^{\prime}_{l\mu}\mathcal{N}_{\mathbf{k}}(\alpha)\Bigg(\sum_{{l^{\prime}}=0}^{l}\sum_{{\mu^{\prime}}=\max\atop {(0,{l^{\prime}}-l+\mu)}}^{\min({l^{\prime}},\mu)} \frac{A_{{l^{\prime}},{\mu^{\prime}}}^{l,\mu}}{N_{{l^{\prime}}}(\alpha)} \Big[\frac{\phi_{l^{\prime}}^{\mu^{\prime}}(\alpha,\mathbf{r}-\mathbf{C}^{\dagger})}{N^\prime_{{l^{\prime}}{\mu^{\prime}}}}z_{l-{l^{\prime}}}^{\mu-{\mu^{\prime}}}\Big(\frac{i\mathbf{k}}{2\alpha}\Big)\nonumber \\
        &&-(1-\delta_{{\mu^{\prime}},0})(1-\delta_{{\mu^{\prime}},\mu})\frac{\phi_{l^{\prime}}^{-{\mu^{\prime}}}(\alpha,\mathbf{r}-\mathbf{C}^{\dagger})}{N^\prime_{{l^{\prime}}-{\mu^{\prime}}}}z_{l-{l^{\prime}}}^{-(\mu-{\mu^{\prime}})}\Big(\frac{i\mathbf{k}}{2\alpha}\Big)\Big]\nonumber \\
        &&+\sum_{{l^{\prime}}=\mu+1}^{l-1}\sum_{{\mu^{\prime}}=\mu+1}^{\min({l^{\prime}},l-{l^{\prime}}+\mu)} \frac{B_{{l^{\prime}},{\mu^{\prime}}}^{l,\mu}}{N_{{l^{\prime}}}(\alpha)} \Big[\frac{\phi_{l^{\prime}}^{\mu^{\prime}}(\alpha,\mathbf{r}-\mathbf{C}^{\dagger})}{N^\prime_{{l^{\prime}}{\mu^{\prime}}}}z_{l-{l^{\prime}}}^{{\mu^{\prime}}-\mu}\Big(\frac{i\mathbf{k}}{2\alpha}\Big)\nonumber \\
        &&+ \frac{\phi_{l^{\prime}}^{-{\mu^{\prime}}}(\alpha,\mathbf{r}-\mathbf{C}^{\dagger})}{N^\prime_{{l^{\prime}}-{\mu^{\prime}}}}z_{l-{l^{\prime}}}^{-({\mu^{\prime}}-\mu)}\Big(\frac{i\mathbf{k}}{2\alpha}\Big)\Big]\nonumber \\
        &&+ \sum_{{l^{\prime}}=1}^{l-\mu-1}\sum_{{\mu^{\prime}}=1}^{\min({l^{\prime}},l-{l^{\prime}}-\mu)} \frac{C_{{l^{\prime}},{\mu^{\prime}}}^{l,\mu}}{N_{{l^{\prime}}}(\alpha)} \Big[\frac{\phi_{l^{\prime}}^{\mu^{\prime}}(\alpha,\mathbf{r}-\mathbf{C}^{\dagger})}{N^\prime_{{l^{\prime}}{\mu^{\prime}}}}z_{l-{l^{\prime}}}^{{\mu^{\prime}}+\mu}\Big(\frac{i\mathbf{k}}{2\alpha}\Big)\nonumber \\
        &&+ \frac{\phi_{l^{\prime}}^{-{\mu^{\prime}}}(\alpha,\mathbf{r}-\mathbf{C}^{\dagger})}{N^\prime_{{l^{\prime}}-{\mu^{\prime}}}}z_{l-{l^{\prime}}}^{-({\mu^{\prime}}+\mu)}\Big(\frac{i\mathbf{k}}{2\alpha}\Big)\Big]\Bigg),
    \end{eqnarray}

    \item for $\mu<0$
    \begin{eqnarray}\label{eqa7}
    e^{i\mathbf{k}\cdot\mathbf{r}}\phi_{l}^{-|\mu|}(\alpha,\mathbf{r}-\mathbf{C}) &=&N_{l}(\alpha)N^{\prime}_{l\mu}\mathcal{N}_{\mathbf{k}}(\alpha)\Bigg(\sum_{{l^{\prime}}=0}^{l} \sum_{{\mu^{\prime}}=\max\atop {(0,{l^{\prime}}-l+|\mu|)}}^{\min({l^{\prime}},|\mu|)} \frac{A_{{l^{\prime}},{\mu^{\prime}}}^{l,|\mu|}}{N_{{l^{\prime}}}(\alpha)} \Big[(1-\delta_{{\mu^{\prime}},|\mu|})\frac{\phi_{l^{\prime}}^{\mu^{\prime}}(\alpha,\mathbf{r}-\mathbf{C}^{\dagger})}{N^\prime_{{l^{\prime}}{\mu^{\prime}}}}\nonumber\\
    &&\times z_{l-{l^{\prime}}}^{-(|\mu|-{\mu^{\prime}})}\Big(\frac{i\mathbf{k}}{2\alpha}\Big) + (1-\delta_{{\mu^{\prime}},0})\frac{\phi_{l^{\prime}}^{-{\mu^{\prime}}}(\alpha,\mathbf{r}-\mathbf{C}^{\dagger})}{N^\prime_{{l^{\prime}}-{\mu^{\prime}}}}z_{l-{l^{\prime}}}^{(|\mu|-{\mu^{\prime}})}\Big(\frac{i\mathbf{k}}{2\alpha}\Big)\Big]\nonumber \\
    &&+ \sum_{{l^{\prime}}=|\mu|+1}^{l-1} \sum_{{\mu^{\prime}}=|\mu|+1}^{\min({l^{\prime}},l-{l^{\prime}}+|\mu|)} \frac{B_{{l^{\prime}},{\mu^{\prime}}}^{l,|\mu|}}{N_{{l^{\prime}}}(\alpha)} \Big[-\frac{\phi_{l^{\prime}}^{\mu^{\prime}}(\alpha,\mathbf{r}-\mathbf{C}^{\dagger})}{N^\prime_{{l^{\prime}}{\mu^{\prime}}}}z_{l-{l^{\prime}}}^{-({\mu^{\prime}}-|\mu|)}(\mathbf{r}_2)\nonumber \\
    && + \frac{\phi_{l^{\prime}}^{-{\mu^{\prime}}}(\alpha,\mathbf{r}-\mathbf{C}^{\dagger})}{N^\prime_{{l^{\prime}}-{\mu^{\prime}}}}z_{l-{l^{\prime}}}^{({\mu^{\prime}}-|\mu|)}\Big(\frac{i\mathbf{k}}{2\alpha}\Big)\Big]\nonumber \\
    &&+ \sum_{{l^{\prime}}=1}^{l-|\mu|-1} \sum_{{\mu^{\prime}}=1}^{\min({l^{\prime}},l-{l^{\prime}}-|\mu|)} \frac{C_{{l^{\prime}},{\mu^{\prime}}}^{l,|\mu|}}{N_{{l^{\prime}}}(\alpha)} \Big[\frac{\phi_{l^{\prime}}^{\mu^{\prime}}(\alpha,\mathbf{r}-\mathbf{C}^{\dagger})}{N^\prime_{{l^{\prime}}{\mu^{\prime}}}}z_{l-{l^{\prime}}}^{-({\mu^{\prime}}+|\mu|)}\Big(\frac{i\mathbf{k}}{2\alpha}\Big)\nonumber \\
    && - \frac{\phi_{l^{\prime}}^{-{\mu^{\prime}}}(\alpha,\mathbf{r}-\mathbf{C}^{\dagger})}{N^\prime_{{l^{\prime}}-{\mu^{\prime}}}}z_{l-{l^{\prime}}}^{(\mu+|\mu|)}\Big(\frac{i\mathbf{k}}{2\alpha}\Big)\Big]\Bigg).
    \end{eqnarray}
\end{enumerate}
Expressions from Eqs.~(\ref{eqa6}) and~(\ref{eqa7}) can be easily transformed into momentum space by Fourier transforming the functions
\(\phi_{l^{\prime}}^{\mu^{\prime}}(\alpha,\mathbf{r}-\mathbf{C}^{\dagger})\),
since all remaining terms are independent of \(\mathbf{r}\).
For convenience, we split the expression in Eq.~\eqref{eqa6} into three terms,
\(a_{(1)}(\alpha,\mathbf{C}^\dagger,\mathbf{k})\),
\(a_{(2)}(\alpha,\mathbf{C}^\dagger,\mathbf{k})\), and
\(a_{(3)}(\alpha,\mathbf{C}^\dagger,\mathbf{k})\),
which are proportional to
\(A_{l^{\prime}\mu^{\prime}}^{l\mu}\),
\(B_{l^{\prime}\mu^{\prime}}^{l\mu}\), and
\(C_{l^{\prime}\mu^{\prime}}^{l\mu}\), respectively.
Similarly, the expression in Eq.~\eqref{eqa7} can be decomposed into the three terms denoted as 
\(b_{(1)}(\alpha,\mathbf{C}^\dagger,\mathbf{k})\),
\(b_{(2)}(\alpha,\mathbf{C}^\dagger,\mathbf{k})\), and
\(b_{(3)}(\alpha,\mathbf{C}^\dagger,\mathbf{k})\), respectively. Thus, the final PW-SGTO matrix elements given in Eq.~\eqref{eq88} can be expressed in four different forms depending on the values of \(\mu_a\) and \(\mu_b\), namely
\begin{enumerate}
    \item for $\mu_a\ge0$ and $\mu_b\ge0$
\begin{equation}\label{eqa8}
O_{l_a \mu_a}^{l_b \mu_b}(\alpha,\beta,\mathbf{R}^{\dagger},\mathbf{k_1},\mathbf{k_2})=D_{l_a\mu_a}^{l_b\mu_b}(\alpha,\beta)\sum_{i,j=1}^{3}\langle a_{(i)}(\alpha,\mathbf{A}^\dagger,\mathbf{k_1})|\widehat{\mathrm{O}}|a_{(j)}(\beta,\mathbf{B}^\dagger,\mathbf{k_2})\rangle,
\end{equation}
    \item for $\mu_a\ge0$ and $\mu_b<0$
\begin{equation}\label{eqa9}
O_{l_a \mu_a}^{l_b \mu_b}(\alpha,\beta,\mathbf{R}^{\dagger},\mathbf{k_1},\mathbf{k_2})=D_{l_a\mu_a}^{l_b\mu_b}(\alpha,\beta)\sum_{i,j=1}^{3}\langle a_{(i)}(\alpha,\mathbf{A}^\dagger,\mathbf{k_1})|\widehat{\mathrm{O}}|b_{(j)}(\beta,\mathbf{B}^\dagger,\mathbf{k_2})\rangle,
\end{equation}
    \item for $\mu_a<0$ and $\mu_b<0$
\begin{equation}\label{eqa10}
O_{l_a \mu_a}^{l_b \mu_b}(\alpha,\beta,\mathbf{R}^{\dagger},\mathbf{k_1},\mathbf{k_2})=D_{l_a\mu_a}^{l_b\mu_b}(\alpha,\beta)\sum_{i,j=1}^{3}\langle b_{(i)}(\alpha,\mathbf{A}^\dagger,\mathbf{k_1})|\widehat{\mathrm{O}}|b_{(j)}(\beta,\mathbf{B}^\dagger,\mathbf{k_2})\rangle,
\end{equation}

    \item for $\mu_a<0$ and $\mu_b\ge0$
\begin{equation}\label{eqa11}
O_{l_a \mu_a}^{l_b \mu_b}(\alpha,\beta,\mathbf{R}^{\dagger},\mathbf{k_1},\mathbf{k_2})=D_{l_a\mu_a}^{l_b\mu_b}(\alpha,\beta)\sum_{i,j=1}^{3}\langle b_{(i)}(\alpha,\mathbf{A}^\dagger,\mathbf{k_1})|\widehat{\mathrm{O}}|a_{(j)}(\beta,\mathbf{B}^\dagger,\mathbf{k_2})\rangle.
\end{equation}
\end{enumerate}
Substituting the expansions from Eqs.~\eqref{eqa6} and~\eqref{eqa7} into the general expression yields the complete matrix-element formulas given in Eqs.~\eqref{eqa8}--\eqref{eqa11}. Each matrix element consists of a sum of nine terms. For case~1, the first term is given by
\begin{eqnarray}\label{eqa12}
    \langle a_{(1)}(\alpha,\mathbf{A}^\dagger,\mathbf{k_1})|\widehat{\mathrm{O}}|a_{(1)}(\beta,\mathbf{B}^\dagger,\mathbf{k_2})\rangle &= & \sum_{{l_a^{\prime}}=0}^{l_a} \sum_{{\mu_a^{\prime}}=\max\atop {(0,{l_a^{\prime}}-l_a+\mu_a)}}^{\min({l_a^{\prime}},\mu_a)} \sum_{{l_b^{\prime}}=0}^{l_b} \sum_{{\mu_b^{\prime}}=\max\atop {(0,{l_b^{\prime}}-l_b+\mu_b)}}^{\min({l_b^{\prime}},\mu_b)} A_{{l_a^{\prime}}{\mu_a^{\prime}}}^{l_a\mu_a} A_{{l_b^{\prime}}{\mu_b^{\prime}}}^{l_b\mu_b}\nonumber \\
    &&\times\frac{1}{N_{{l_a^{\prime}}{l_b^{\prime}}}(\alpha,\beta)} \Bigg[ \frac{O_{{l_a^{\prime}}{\mu_a^{\prime}}}^{\prime{l_b^{\prime}}{\mu_b^{\prime}}}(\alpha,\beta,\mathbf{R}^{\dagger},\mathbf{k_1},\mathbf{k_2})}{N_{{l_a^{\prime}}{\mu_a^{\prime}}}^{\prime{l_b^{\prime}}{\mu_b^{\prime}}}}\nonumber\\
    &&\times z_{l_a-{l_a^{\prime}}}^{\mu_a-{\mu_a^{\prime}}} \Big(-\frac{i\mathbf{k_1}}{2\alpha}\Big) z_{l_b-{l_b^{\prime}}}^{\mu_b-{\mu_b^{\prime}}} \Big(\frac{i\mathbf{k_2}}{2\beta}\Big)\nonumber \\
    && - (1-\delta_{{\mu_b^{\prime}},0})(1-\delta_{{\mu_b^{\prime}},\mu_b}) \frac{O_{{l_a^{\prime}}{\mu_a^{\prime}}}^{\prime{l_b^{\prime}}-{\mu_b^{\prime}}}(\alpha,\beta,\mathbf{R}^{\dagger},\mathbf{k_1},\mathbf{k_2})}{N_{{l_a^{\prime}}{\mu_a^{\prime}}}^{\prime{l_b^{\prime}}-{\mu_b^{\prime}}}}\nonumber \\
    &&\times z_{l_a-{l_a^{\prime}}}^{\mu_a-{\mu_a^{\prime}}} \Big(-\frac{i\mathbf{k_1}}{2\alpha}\Big) z_{l_b-{l_b^{\prime}}}^{-(\mu_b-{\mu_b^{\prime}})} \Big(\frac{i\mathbf{k_2}}{2\beta}\Big)\nonumber \\
    && -(1-\delta_{{\mu_a^{\prime}},0})(1-\delta_{{\mu_a^{\prime}},\mu_a}) \frac{O_{{l_a^{\prime}}-{\mu_a^{\prime}}}^{\prime{l_b^{\prime}}{\mu_b^{\prime}}}(\alpha,\beta,\mathbf{R}^{\dagger},\mathbf{k_1},\mathbf{k_2})}{N_{{l_a^{\prime}}-{\mu_a^{\prime}}}^{\prime{l_b^{\prime}}{\mu_b^{\prime}}}}\nonumber \\
    &&\times z_{l_a-{l_a^{\prime}}}^{-(\mu_a-{\mu_a^{\prime}})} \Big(-\frac{i\mathbf{k_1}}{2\alpha}\Big) z_{l_b-{l_b^{\prime}}}^{\mu_b-{\mu_b^{\prime}}} \Big(\frac{i\mathbf{k_2}}{2\beta}\Big)\nonumber \\
    &&+ \frac{O_{{l_a^{\prime}}-{\mu_a^{\prime}}}^{\prime{l_b^{\prime}}-{\mu_b^{\prime}}}(\alpha,\beta,\mathbf{R}^{\dagger},\mathbf{k_1},\mathbf{k_2})}{N_{{l_a^{\prime}}-{\mu_a^{\prime}}}^{\prime{l_b^{\prime}}-{\mu_b^{\prime}}}}\nonumber\\
    &&\times z_{l_a-\lambda_a}^{-(\mu_a-{\mu_a^{\prime}})} \Big(-\frac{i\mathbf{k_1}}{2\alpha}\Big) z_{l_b-{l_b^{\prime}}}^{-(\mu_b-{\mu_b^{\prime}})} \Big(\frac{i\mathbf{k_2}}{2\beta}\Big) \Bigg]
\end{eqnarray}
The remaining terms can be obtained analogously.

\twocolumngrid


%

\end{document}